**Moderate D/H ratios in methane ice on Eris and Makemake as evidence of hydrothermal or metamorphic processes in their interiors: Geochemical analysis**

Christopher R. Glein [1*], William M. Grundy [2,3], Jonathan I. Lunine [4], Ian Wong [5,6], Silvia Protopapa [7], Noemi Pinilla-Alonso [8], John A. Stansberry [9,2,3], Bryan J. Holler [9], Jason C. Cook [10], Ana Carolina Souza-Feliciano [8]

[1] Space Science Division, Space Sector, Southwest Research Institute, 6220 Culebra Road, San Antonio, TX 78238-5166, United States

[2] Lowell Observatory, Flagstaff, AZ

[3] Department of Astronomy and Planetary Science, Northern Arizona University, Flagstaff, AZ

[4] Department of Astronomy and Carl Sagan Institute, Cornell University, Ithaca, NY

[5] NASA Goddard Space Flight Center, Greenbelt, MD

[6] Department of Physics, American University, Washington, DC

[7] Southwest Research Institute, Boulder, CO

[8] Florida Space Institute, University of Central Florida, Orlando, FL

[9] Space Telescope Science Institute, Baltimore, MD

[10] Pinhead Institute, Telluride, CO

[*] Corresponding author. Email: christopher.glein@swri.org

**Highlights**

- Models of D/H fractionation can constrain the origin of methane on Eris and Makemake.
- Primordial methane would have a D/H ratio that disagrees with data from JWST.
- D/H constraints point to methane production inside of these worlds.
- Elevated temperatures in rocky cores are implied to produce methane.
- Future observations may indicate the extent and duration of endogenic activity.

**Abstract**

Dwarf planets Eris and Makemake have surfaces bearing methane ice of unknown origin. This ice can provide important insights into the origin and evolution of volatiles in the outer solar system. Deuterium/hydrogen (D/H) ratios were recently determined from *James Webb Space Telescope* (JWST) observations of Eris and Makemake (Grundy et al., 2024b), giving us new clues to decipher the origin of methane. Here, we develop geochemical models to test if the origin of methane could be primordial, derived from $CO_2$ or CO ("abiotic"), or sourced by organics ("thermogenic"). We find that primordial methane (as currently understood) is inconsistent with the observational data, whereas both abiotic and thermogenic methane can have D/H ratios that overlap the observed ranges. This suggests that Eris and Makemake either never acquired much methane during their formation, or their original inventories were removed and then replaced by internally produced methane. Because producing abiotic or thermogenic methane likely requires temperatures above ~150°C, we infer that Eris and Makemake have rocky cores that underwent substantial radiogenic heating. Their cores may still be warm/hot enough to make methane. This heating could have driven hydrothermal circulation at the bottom of an ice-covered ocean to generate abiotic methane, and/or metamorphic reactions involving accreted organic matter could have occurred in response to heating in the deeper interior, generating thermogenic methane. Additional analyses of relevant thermal evolution model results and theoretical predictions of the D/H ratio of methane in the solar nebula support our findings of elevated subsurface temperatures and an apparent lack of primordial methane on Eris and Makemake. It remains an open question whether their D/H ratios may have evolved subsequent to methane outgassing. We also suggest that lower-than-expected D/H and $^{84}Kr/CH_4$ ratios in Titan's atmosphere disfavor a primordial origin of methane there as well. Recommendations are given for future activities to further test proposed scenarios of abiotic and thermogenic methane production on Eris and Makemake, and to explore these worlds up close so that we can see if they bear additional evidence of endogenic processes.

# 1. Introduction

The surfaces of large trans-Neptunian objects (TNOs) hold clues to how icy worlds received their volatile endowments. Such clues are needed to understand formation conditions of planetary building blocks in the early outer solar system (e.g., Schneeberger et al., 2023), and the thermal and geodynamic evolution of icy worlds (e.g., Loveless et al., 2022). By understanding these phenomena, we can gain new insights into the origin and evolution of the solar system, as well as broaden our perspective of the habitability potential of icy worlds. These are major goals of NASA and the planetary science and astrobiology communities (NASEM, 2022).

Dwarf planets Eris and Makemake are alluring as they are two of the largest TNOs. Eris is nearly the size of Pluto (Sicardy et al., 2011), and Makemake is larger than Pluto's moon Charon (Ortiz et al., 2012; Brown, 2013) (see the companion paper by Grundy et al., 2024b for more comprehensive information on these bodies). Because they are big and their surfaces cold, volatiles exist as stable ice deposits (Schaller and Brown, 2007; Brown et al., 2011; Johnson et al., 2015), allowing us to probe their chemical nature and infer what they reveal about the history of the body. Subsolar temperatures at the surfaces of these bodies have been estimated to range from ~30 to ~40 K (Sicardy et al., 2011; Ortiz et al., 2012), which can be compared to the observed range of ~37-55 K on Pluto (Hinson et al., 2017). Eris and Makemake have relatively high bulk densities (Eris, ~2400 kg/m$^3$, Holler et al., 2021; Makemake, ~1700-2100 kg/m$^3$, Parker et al., 2018), which provide clues to their internal evolution, as discussed later (see Section 3.2). It has been known for many years that Eris and Makemake have surfaces that are dominated spectrally by methane ice (Brown et al., 2005, 2007, 2015; Licandro et al., 2006a, 2006b; Dumas et al., 2007; Tegler et al., 2008, 2010, 2012; Merlin et al., 2009; Alvarez-Candal et al., 2011, 2020; Lorenzi et al., 2015; Perna et al., 2017). This finding was recently corroborated using near-infrared spectra from the *James Webb Space Telescope* (JWST) (Grundy et al., 2024b).

Most spectacularly, JWST discovered monodeuterated methane (CH$_3$D) on the surfaces of Eris and Makemake, and deuterium/hydrogen (D/H) ratios in methane were determined (Eris = (2.5±0.5)×10$^{-4}$, Makemake = (2.9±0.6)×10$^{-4}$, 1σ; Grundy et al., 2024b). These D/H data provide an opportunity to obtain new constraints on the origin of methane on large TNOs. Isotopic data are more diagnostic of provenance than the molecular composition is, because many processes can supply the same species but they typically give different isotopic ratios (e.g., Schoell, 1988). Indeed, previous work on Titan's methane demonstrated that the D/H ratio can serve as a powerful proxy of origin and evolution (Owen et al., 1986; Pinto et al., 1986; Lunine et al., 1999; Cordier et al., 2008; Mandt et al., 2009, 2012; Nixon et al., 2012).

Here, we aim to infer what the observed D/H ratios may mean for how Eris and Makemake acquired methane, and to show how this knowledge helps to illuminate the poorly understood formation conditions and internal evolution of these worlds. This paper is a D/H geochemistry-focused companion to Grundy et al. (2024b), who focus on the JWST data and complement our D/H interpretation with more discussion on the implications of $^{13}$C/$^{12}$C ratios, the non-detection of CO on both bodies, and the lack of detectable N$_2$ on Makemake. Section 2 presents a model for the D/H ratios of methane derived from different origin scenarios. Based on the Titan literature (Mousis et al., 2009a; Glein, 2015; Miller et al., 2019), there are three types of methane most likely to exist on Eris and Makemake: (1) primordial methane that was present in the protoplanetary disk or interstellar medium, (2) so-called abiotic methane produced by hydrothermal processing of carbon dioxide or carbon monoxide in the presence of rocky materials, and (3) thermogenic methane derived from complex organic molecules (see Section 2 for details). We do not pursue investigation of biologically produced methane because non-biological processes can explain the data, and life is the hypothesis of last resort (Sagan et al., 1993; Neveu et al., 2018). Section 3 presents D/H predictions and comparisons with observed D/H ratios to assess whether Eris's and Makemake's methane could be primordial, abiotic, or thermogenic. We discuss implications for the formation conditions and internal evolution of the two bodies, as well as current uncertainties. As a point of comparison, we also discuss how our new understanding of D/H ratios and noble gas abundances may clarify the origin of methane on Titan. We conclude this paper in Section 4, where we provide a summary of our findings and recommendations for future work.

## 2. Geochemical model

*2.1. Modeling philosophy*

We seek to develop a framework that enables predictions of the D/H ratio of surface methane if it were derived from different hypothesized sources. To proceed, we need to find ways to predict the D/H ratio of methane in terms of quantities that can be constrained for Eris and Makemake. Here, we construct simple models of isotopic fractionation in an attempt at making this connection. The driving compositional parameter in these models is the isotope fractionation factor, which can be written in generic form as

$$\alpha_{i\text{-}j} = \frac{\left(\frac{D}{H}\right)_i}{\left(\frac{D}{H}\right)_j}, \qquad (1)$$

where *i* and *j* correspond to different compounds or reservoirs containing hydrogen atoms. Our general strategy is to rely on empirical data. The rationale for this is that many of the processes that can provide methane on planetary bodies are not well-understood (Atreya, 2007; Reeves and Fiebig, 2020; Thompson et al., 2022); it makes little sense to attempt to develop detailed theoretical models at this stage. Hence, our model will not explicitly include dependences on time, temperature, reaction networks, and other kinetically relevant variables. Instead, we will adopt fractionation factors from direct measurements of different types of methane. This is a practical decision that makes sense for a first interpretation of the JWST data; however, we should keep in mind that our approach does not allow assessment of parts of the parameter space that may be theoretically allowed but are not presently known to be represented by observations. In our approach, we treat the D/H ratio of water as the master variable (see Section 3.1). This is a convenient choice since there are a number of observational data on the D/H ratio of cometary water (see Biver et al., in press). Some comets (i.e., Jupiter-family comets) are probably similar to the building blocks of Eris and Makemake (Fraser et al., in press).

*2.2. Primordial methane*

Primordial methane could have been delivered to Eris and Makemake directly as methane. Exogenous and endogenous sources can be envisioned. In the former case, comets (e.g., Dello Russo et al., 2016) would have brought methane to the surfaces of these bodies. The D/H ratio of methane has been measured in only one comet, 67P/Churyumov-Gerasimenko (Müller et al., 2022) (note: a tentative determination of the D/H ratio of methane was reported for another comet; see Section 3.3). The D/H ratio of methane in comet 67P could be adopted as a single value for all comets, but this may be hasty given that we already know that water shows variability in its D/H ratio among comets (Biver et al., in press). We see no reason why methane should be different. Therefore, we opt to take a more general approach by allowing the D/H ratios of methane and water to vary, but we assume the fractionation factor between them to be roughly constant. We then estimate the D/H ratio of methane delivered by comets via

$$\left(\frac{D}{H}\right)^{primo,exo}_{CH_4,surf} = E_{esc/chem}\, \alpha^{OSN}_{CH_4\text{-}H_2O} \times \left(\frac{D}{H}\right)_{H_2O}. \tag{2}$$

All model parameters introduced in Section 2 are defined in Table 1. Figure 1 shows how these parameters relate to processes that govern the D/H ratio of methane during the evolution of Eris/Makemake, for four candidate types of methane. The D/H fractionation factor between $CH_4$ and $H_2O$ in the outer solar nebula, denoted by $\alpha_{CH4\text{-}H2O}^{OSN}$, is critical for constraining the D/H ratio of primordial methane. Equation 2 also includes a factor ($E_{esc/chem}$) that accounts for deuterium enrichment that occurs due to preferential loss of $CH_4$ (relative to $CH_3D$), depending on the age of the surface inventories of methane on Eris and Makemake. If their inventories are young (geologically speaking), this factor could be close to unity (see Grundy et al., 2024b). In our calculations, we assume this factor to be unity since the actual value may not be much larger, as discussed in Section 3.4.

**Table 1.** Model parameters for predicting D/H ratios of methane on icy worlds.

| Parameter | Symbol | Value [a] | References |
|---|---|---|---|
| D/H enrichment factor from atmospheric evolution | $E_{esc/chem}$ | ≥1 | See Section 3.4 |
| D/H fractionation factor between $CH_4$ and $H_2O$ in the outer solar nebula | $\alpha^{OSN}_{CH_4\text{-}H_2O}$ | 4.1-5.5 | Müller et al. (2022) |
| D/H fractionation factor between $CH_4$ vapor and hydrate during cryovolcanic outgassing | $\alpha^{CH_4}_{vap\text{-}hyd}$ | 1.01 | Hachikubo et al. (2023) |
| D/H fractionation factor between abiotic $CH_4$ and $H_2O$ | $\alpha^{abio}_{CH_4\text{-}H_2O}$ | 0.60-0.90 | Wang et al. (2018); Warr et al. (2021a, 2021b) |
| Molar ratio of carbon as accreted organic matter to silicon | $\frac{C_{OM}}{Si}$ | 0.77-6 | Alexander et al. (2017a); Bardyn et al. (2017) |
| Molar ratio of hydrogen to carbon in accreted organic matter | $\left(\frac{H}{C}\right)_{OM}$ | 0.8-1.2 | Kissel and Krueger (1987); Alexander et al. (2007); Isnard et al. (2019) |
| Molar ratio of equivalent water in phyllosilicates to silicon in rock after differentiation | $\frac{(H_2O)_{phyllo}}{Si}$ | 1.78 | Alexander (2019) |
| D/H depletion factor of organic matter from H-D exchange during water-rock differentiation | $D_{alt}$ | 0.39-0.78 | Alexander et al. (2007); Piani et al. (2021); see Section 2.4 |
| D/H fractionation factor between organic matter and $H_2O$ in the outer solar nebula | $\alpha^{OSN}_{OM\text{-}H_2O}$ | 3.7 | Alexander et al. (2007, 2012); Paquette et al. (2021); Müller et al. (2022); see Section 2.4 |
| D/H fractionation factor between phyllosilicates and $H_2O$ during water-rock differentiation | $\alpha^{diff}_{phyllo\text{-}H_2O}$ | 0.94 | Saccocia et al. (2009) |
| D/H fractionation factor between thermogenic $CH_4$ and $H_2O$ | $\alpha^{thermo}_{CH_4\text{-}H_2O}$ | 0.80-0.89 | Wang et al. (2015); Giunta et al. (2019) |

[a], Ranges are 1σ, or what we find to be the most concordant range when results from multiple studies are integrated. See Section 2 for the context on TNOs.

**Figure 1.** Stages in the origin and evolution of Eris and Makemake, with a focus on materials (written inside boxes) and chemical/physical processes (indicated in red) that are hypothesized to explain the presence of methane. Each row is distinguished by methane from a source type named in bold, black text. Model parameters (see Table 1) that represent the isotopic consequences of the corresponding process are shown in blue. Other materials or processes could occur, but those shown here are most significant to the D/H ratio of methane.

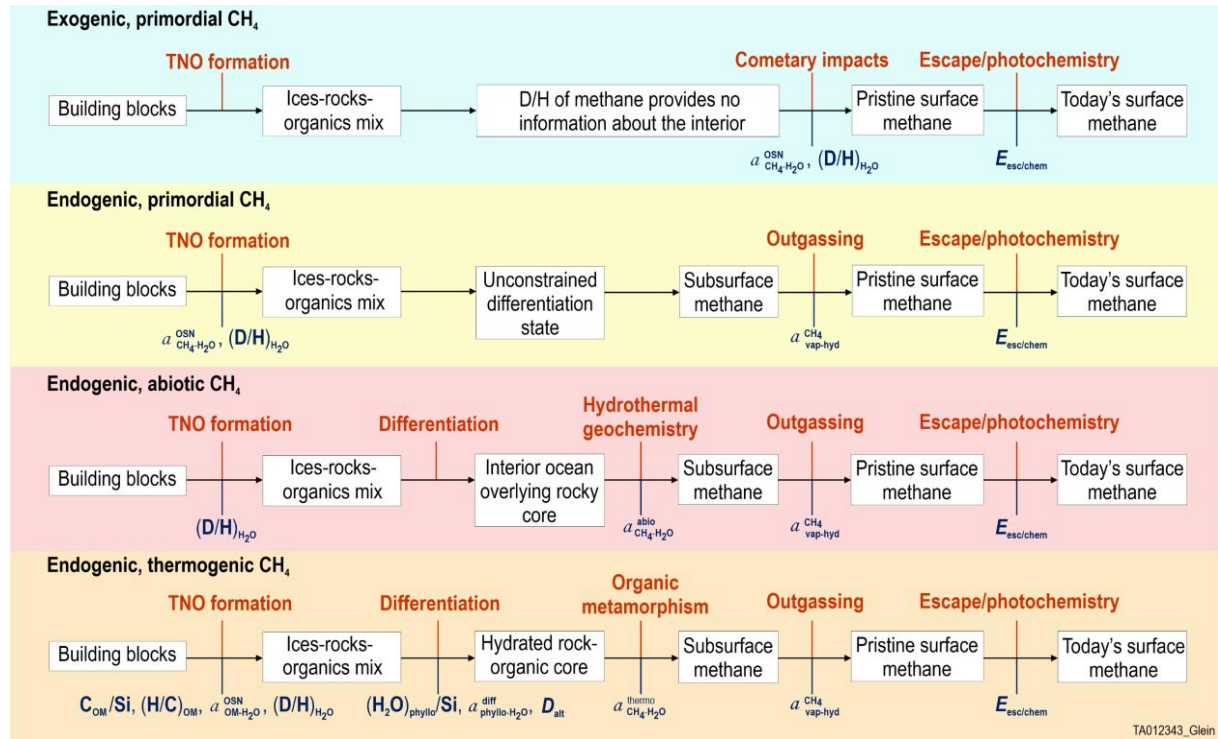

An alternative source of primordial methane is accretion of methane-bearing ices during the formation of Eris and Makemake, followed by outgassing of methane from their interiors. Features indicative of cryovolcanism, including methane outgassing, have been identified on Pluto (Singer et al., 2022; Howard et al., 2023). By analogy, related processes might occur on Eris and Makemake (Grundy, 2020). We need to understand if outgassing might impart fractionation. Because only phase transitions of methane are involved, we do not expect a large degree of isotopic fractionation, but this needs to be verified. Assuming that Eris and Makemake differentiated into a rocky core and a hydrosphere (see Section 3.2), the formation of clathrate hydrates is likely to occur during methane transport through the icy shell (e.g., Kamata et al., 2019). Since clathrates can trap large amounts of methane (up to 1 mol methane per 5.75 mol water), we assume that they are the dominant source of methane outgassing from the subsurface (e.g., Howard et al., 2023). The isotope effect from clathrate formation can be accounted for by incorporating an additional fractionation factor into the previous equation, which yields

$$\left(\frac{D}{H}\right)_{CH_4,surf}^{primo,endo} = E_{esc/chem}\, \alpha_{vap\text{-}hyd}^{CH_4}\, \alpha_{CH_4\text{-}H_2O}^{OSN} \times \left(\frac{D}{H}\right)_{H_2O} \qquad (3)$$

for primordial methane that is endogenic. Experimental data (Hachikubo et al., 2023) show that clathrates barely fractionate $CH_3D$ from $CH_4$; the fractionation factor is close to unity (see Table 1).

Our nominal assumption is to use a constant value of $\alpha_{CH4\text{-}H2O}^{OSN}$ based on comet 67P (see Table 1). This is the simplest case consistent with observations, so we consider this a reasonable place to start. However, we recognize that it may turn out to be too simplistic to assume that a uniform fractionation factor is representative of the outer solar nebula. Therefore, it is important to understand how nebular processes could have affected the fractionation factor. We explore the effects of two processes later in this work – mixing of interstellar-derived and solar-equilibrated isotopic reservoirs of methane and water (see the Appendix), and D loss from methane and water whose original interstellar D enrichments (Sakai et al., 2012; Albertsson et al., 2013) are gradually erased via isotopic exchange with $H_2$ (see Section 3.3). By "interstellar", we mean methane formed in cold (<25 K) molecular clouds (Qasim et al., 2020). These cases provide pathways to producing D/H ratios in between interstellar and protosolar values. One case relies on physical processes, the other on chemical processes. Both can provide insights into how the fractionation factor might differ from assumed values.

*2.3. Abiotic methane*

Methane could be produced as a consequence of water-rock reactions (Glein et al., 2008; Etiope and Sherwood Lollar, 2013) below the seafloors of Eris and Makemake, if these worlds have (or had) interior water oceans (Hussmann et al., 2006). For comparison, Guo and Eiler (2007) proposed that methane was synthesized on some carbonaceous chondrite parent bodies and could have been synthesized on some Kuiper belt objects. In fact, methane is found in some carbonaceous chondrites; it may have an abiotic origin (Glein and Zolotov, 2020). The source of carbon would be accreted $CO_2$ or CO, similar to what we find in comets (Biver et al., in press). Molecular hydrogen derived from water-rock reactions (e.g., Vance et al., 2007; Waite et al., 2017) could support the synthesis of methane if transition metal-bearing catalysts are present (Horita and Berndt, 1999; McGlynn et al., 2020). The D/H ratio of such abiotic methane can be calculated using the equation shown below

$$\left(\tfrac{D}{H}\right)^{abio,endo}_{CH_4,surf} = E_{esc/chem}\, \alpha^{CH_4}_{vap\text{-}hyd}\, \alpha^{abio}_{CH_4\text{-}H_2O} \times \left(\tfrac{D}{H}\right)_{H_2O}. \qquad (4)$$

We can set broad limits on $\alpha^{abio}_{CH4\text{-}H2O}$ by examining the δD ranges [1] of abiotic methane on Earth as conventionally defined by Milkov and Etiope (2018). From that investigation, we found that the most D-rich abiotic methane is from submarine hydrothermal fluids, and the least D-rich abiotic methane is from fracture fluids hosted in Precambrian crystalline rocks. Isotopic data from these environments (Wang et al., 2018; Warr et al., 2021a, 2021b) were used to constrain the fractionation factor for abiotic methane (see Table 1).

*2.4. Thermogenic methane*

Thermogenic methane is that derived from the decomposition of organic materials caused by heating of source rocks. The carbon comes from organic carbon (it is important to note that, while we contrast this process with what we call "abiotic methane", we are not implying that thermogenesis is acting upon organics of biological origins). Hydrogen atoms can be supplied by organic matter, but also by hydrated minerals such as phyllosilicates and other sources of water (Schimmelmann et al., 2006). It can be expected that if Eris and Makemake formed at warm enough temperatures, then CO, primordial $CH_4$, and $CO_2$ would not be present in accreted solids, while more refractory (possibly interstellar) organics could have been incorporated into planetary building blocks. Or, more volatile forms of carbon might have been driven off during or not long after the formation of these bodies (see Section 3.3), leaving organic matter as the main carbon source. For modeling of isotopic fractionation, we may not just take data for thermogenic methane on Earth and apply them directly to Eris and Makemake, because organic matter on Earth is overwhelmingly derived from biological activity (Tissot and Welte, 1984). The organic and inorganic contributions of hydrogen atoms would very likely differ on Eris and Makemake, since the compositions of their presumed cores and sedimentary systems on Earth would be different. So, we must develop a model that is more complicated.

---

[1] In delta notation, the D/H ratio of a sample is expressed relative to that of a standard, which is defined by isotope geochemists as Vienna Standard Mean Ocean Water (VSMOW), with D/H = 1.5576×10$^{-4}$ (Hagemann et al., 1970). The δD range for Eris/Makemake methane is +280 to +1250‰ (1σ).

The first step is to estimate the D/H ratio of the bulk core where thermogenic methane may be generated. This can be done by performing calculations for a mixture of organic matter and phyllosilicates – the most abundant carriers of hydrogen atoms in carbonaceous chondrites that came from water-rich parent bodies (Russell et al., 2022). Comets are not generally recognized to contain phyllosilicates (e.g., Brownlee et al., 2012), but it can be assumed that phyllosilicates would have formed during water-rock differentiation on Eris and Makemake because the timescale of silicate mineral hydration is geologically short (Zandanel et al., 2022). Here, we assume complete hydration (if the initial core was only partially hydrated, then our model would give lower limits on the D/H ratio of thermogenic methane); the existence of low-density (<3000 kg/m$^3$) cores inside Enceladus (Iess et al., 2014) and Titan (Durante et al., 2019) may support this assumption. The mixing equation for hydrogen isotopes is

$$\left(\frac{D}{H}\right)_{\text{bulk core}} = \frac{\frac{C_{OM}}{Si} \times \left(\frac{H}{C}\right)_{OM} \times \left(\frac{D}{H}\right)_{OM} + 2 \times \frac{(H_2O)_{\text{phyllo}}}{Si} \times \left(\frac{D}{H}\right)_{\text{phyllo}}}{\frac{C_{OM}}{Si} \times \left(\frac{H}{C}\right)_{OM} + 2 \times \frac{(H_2O)_{\text{phyllo}}}{Si}}. \tag{5}$$

It is also assumed that icy pebbles/planetesimals that formed beyond Saturn (Hopp et al., 2022) had an organic matter content between that in CI chondrites and comet 67P, and the H/C ratio of accreted organic matter was similar to that of the most primitive organic matter in carbonaceous chondrites and comets (see Table 1). The D/H ratios of organic matter and phyllosilicates can be expressed as

$$\left(\frac{D}{H}\right)_{OM} = D_{\text{alt}} \alpha_{\text{OM-H}_2\text{O}}^{\text{OSN}} \times \left(\frac{D}{H}\right)_{H_2O} \tag{6}$$

and

$$\left(\frac{D}{H}\right)_{\text{phyllo}} = \alpha_{\text{phyllo-H}_2\text{O}}^{\text{diff}} \times \left(\frac{D}{H}\right)_{H_2O}, \tag{7}$$

respectively (see Table 1). These equations are meant to describe the core's D/H composition immediately following a period of differentiation.

Two effects are embedded within Equation 6. One parameter ($\alpha_{\text{OM-H2O}}^{\text{OSN}}$) accounts for D/H fractionation between organic matter and water in the outer solar nebula prior to parent body formation. This organic material is D-rich (see below). The second effect results from isotopic exchange between organic-bound H and water during ice melting. Parameter $D_{\text{alt}}$ quantifies deuterium removal from organic matter in the interior. Just like minerals, organic matter can be subject to aqueous alteration. Affected organic matter will have D/H ratios lower than the original value (see below). Aqueously altered organic material is what becomes incorporated into the cores of Eris and Makemake.

Two endmember scenarios have been proposed to explain the D/H ratios of organic matter in carbonaceous chondrites. One assumes that chondrite parent bodies accreted organic matter whose D/H ratio was determined by where and when accretion occurred (Piani et al., 2021). In this scenario, each parent body accreted organic matter with a distinct D/H ratio. The other endmember assumes that chondrite parent bodies accreted organic matter with a common D/H ratio, and variable H-D exchange between organic matter and water occurred during aqueous alteration (Alexander et al., 2017b). The latter scenario conflicts with the finding that organic matter in comet 67P (~16×10$^{-4}$; Paquette et al., 2021) has a significantly higher D/H ratio than organic matter in CR chondrites (~7×10$^{-4}$; Alexander et al., 2007). On the other hand, it is implausible to assume that H-D exchange would not have occurred in the most aqueously altered carbonaceous chondrites (e.g., CI chondrites), as experiments show that H-D exchange between organic matter and water is relatively fast (Foustoukos et al., 2021; Kebukawa et al., 2021).

Both scenarios may provide pieces of the puzzle. Since CR chondrites and comet 67P have the least altered organic matter, they can be used to characterize the nature of accreted material. While organic matter from these objects has distinct D/H ratios, their values of $\alpha_{OM-H_2O}^{OSN}$ are consistent (CRs = 3.3 to 4.1, 67P = 2.0 to 4.2; Alexander et al., 2007; 2012; Paquette et al., 2021; Müller et al., 2022), assuming $D_{alt} \approx 1$ (no modification). To first-order, this suggests that the fractionation factor can be assumed to be constant, even though individual D/H ratios can vary. We solve Equation 6 to estimate $D_{alt}$ for CI and CM chondrites based on their observed $\alpha_{OM-H_2O}$ (CIs = 1.7 to 1.9, CMs = 2.5 to 2.9; Alexander et al., 2007; Piani et al., 2021) and $\alpha_{OM-H_2O}^{OSN}$ = 3.7. For this case, we find $D_{alt}$ values of 0.46-0.51 and 0.68-0.78, respectively. We adopt a representative value of $\alpha_{OM-H_2O}^{OSN}$ to avoid confusing the ranges of two parameters that are interdependent. This approach does not impact the product of these parameters, which is the quantity that is important for the present application (see Equation 6).

For Eris and Makemake, we consider a range for $D_{alt}$ of 0.39-0.78. This includes the observations of CI and CM chondrites and is extended down to account for the possibility of a greater extent of H-D exchange during the differentiation of Eris and Makemake. If the CI parent body (see McSween et al., 2018) was smaller than Eris and Makemake, there may be a longer duration of water-organic interactions on the latter bodies. Our lower value should be seen as an initial attempt to be more conservative rather than as a hard limit.

The D/H fractionation factor between phyllosilicates and water during water-rock differentiation can be represented by that of the serpentine-water system at 0°C (Saccocia et al., 2009). Serpentine is frequently the most abundant hydrated mineral in aqueously altered chondrites (Brearley, 2006). However, this choice is not critical as the magnitude of isotopic fractionation between hydroxyl (OH) in minerals and water is small ($\alpha_{phyllo-H_2O}^{diff}$ may deviate from the value for serpentine by <5%; see Saccocia et al., 2009).

To simplify calculations of the D/H ratio of thermogenic methane on Eris and Makemake, two endmembers are considered: metamorphic fluids (e.g., Melwani Daswani et al., 2021) that are rich in methane or water (McKinnon et al., 2021). For methane-rich fluids, (D/H)$_{CH_4,core}$ ≈ (D/H)$_{bulk\ core}$. For water-rich fluids, (D/H)$_{H_2O,core}$ ≈ (D/H)$_{bulk\ core}$ and the D/H ratio of methane in the core can be calculated using the following relationship

$$\left(\frac{D}{H}\right)_{CH_4,core} = \alpha_{CH_4-H_2O}^{thermo} \times \left(\frac{D}{H}\right)_{H_2O,core}. \tag{8}$$

The fractionation factor for thermogenic methane in water-rich fluids can be constrained using isotopic data from gas-producing sedimentary systems on Earth (Wang et al., 2015; Giunta et al., 2019). The adopted range of $\alpha_{CH_4-H_2O}^{thermo}$ in Table 1 accounts for any kinetic effects and the approach to isotopic equilibrium (Turner et al., 2021) in Earth analogue systems of warm environments where water, rock, and organics may interact inside large TNOs. It does not account for unfamiliar chemistry that could be hypothesized. Finally, for both endmembers of core fluids, the D/H ratio of methane on the surfaces of Eris and Makemake would be

$$\left(\frac{D}{H}\right)_{CH_4,surf}^{thermo,endo} = E_{esc/chem} \alpha_{vap-hyd}^{CH_4} \times \left(\frac{D}{H}\right)_{CH_4,core}. \tag{9}$$

*2.5. Impact-generated methane?*

One might wonder whether there could be abiotic or thermogenic methane that is derived from exogenic sources on Eris and Makemake. These types of methane would be distinct from intact methane delivered by comets, which falls under the umbrella of primordial methane (see Section 2.2). We can envision that methane might be produced during the impacts of comets. However, cometary impacts are relatively gentle in the trans-Neptunian region because of low (≲4 km/s) impact velocities (Dell'Oro et al., 2013). They also appear to be relatively infrequent (e.g., Singer et al., 2019). It is likely that methane-forming reactions will be kinetically inhibited during impacts on TNOs, as there is not enough energy to generate high temperatures over the short timescales of shock heating (<1 min; Steckloff et al., 2023). As an example, Sekine et al. (2014) experimentally showed that impacts can produce methane from methanol, but only if the impact velocity is greater than ~6 km/s. Even if a statistically unlikely impact were to be sufficiently energetic, the experiments of Sekine et al. (2014) also showed that once the velocity is high enough to produce volatiles, CO rather than $CH_4$ is the dominant carbon species. Modeling studies indicate that this type of speciation is a general consequence of impact chemistry (e.g., Ishimaru et al., 2010). Compositions with more CO than $CH_4$ are inconsistent with the surface compositions of Eris and Makemake (see Grundy et al., 2024b).

## 3. Results and discussion

*3.1. From D/H ratio to source of methane*

With our modeling framework (see Section 2), the D/H ratio of methane can be predicted as a function of the D/H ratio of water. Unfortunately, the D/H ratio of Eris's and Makemake's water is unknown, and water ice has not been observed on their surfaces (Barucci et al., 2011). Nevertheless, water is expected to be a component of their interiors (McKinnon et al., 2008), based on the fact that the bulk densities of Eris (~2400 kg/m$^3$; Holler et al., 2021) and Makemake (~1700-2100 kg/m$^3$; Parker et al., 2018) are between typical values for water (~920-1000 kg/m$^3$) and rock (~2500-3500 kg/m$^3$). Because Jupiter-family comets (JFCs) are thought to have mainly originated in the scattered disk of the Kuiper belt (Fraser et al., in press), we can use D/H measurements of water outgassed from JFCs to determine a reasonable range for Eris and Makemake.

Müller et al. (2022) compiled an extensive amount of D/H data for comets, and from their compilation, we found that comets Hartley 2 (Hartogh et al., 2011) and 67P (Müller et al., 2022) have the most precisely measured D/H ratios of water that span the widest range for JFCs. Accordingly, we adopt the range (1.37-5.41)×10$^{-4}$ to represent water inside Eris and Makemake; this includes 1σ error bars. Note that water outgassed from some comets could differ from water that was originally accreted in the nucleus. Lis et al. (2019) suggested that all comets may share the same Earth-like D/H ratio in accreted water ice. Observed differences in D/H among comets could then be due to different contributions of water outgassed from the surface of the nucleus (high D/H) and from icy grains ejected from deeper in the nucleus (terrestrial D/H). Alternatively, different D/H ratios may reflect different formation locations of comets (Yang et al., 2013). We decided to adopt a wide range of D/H ratios (which also includes Earth's ocean water, and structural water contained in minerals in CI chondrites and asteroid Ryugu samples; Piani et al., 2023) in an effort to be conservative. Also, it seems more consistent to assume isotopic heterogeneity among comets given their known molecular diversity (Biver et al., in press).

Figure 2 shows how the D/H ratio of methane is predicted to relate to the D/H ratio of water for different origins of methane on Eris and Makemake. We begin with this figure as it provides a general survey of our parameter space. Although it may not be clear in the semi-log plot, predicted D/H ratios for all three possible source types are assumed to be proportional to that of water based on the adopted modeling approach. It can be seen that primordial, abiotic, and thermogenic methane occupy largely distinct zones. For a given D/H ratio in water, we can expect primordial methane to have the highest D/H ratios, followed by thermogenic methane, and then abiotic methane. Primordial methane has the highest values because it most likely originated in interstellar environments at very low temperatures that promote D enrichment (Sakai et al., 2012; Albertsson et al., 2013). Abiotic methane has the lowest values because it would inherit hydrogen atoms from water (with a lower D/H ratio) during its synthesis in hydrothermal systems. Thermogenic methane is in the middle because it would obtain hydrogens from a mixture of D-enriched organic matter and phyllosilicates containing water-derived hydrogen (see Section 2.4). The thermogenic zone is the biggest as there is large uncertainty in the organic matter content of accreted rock (see Table 1). Because we cannot tightly constrain the D/H ratio of water on Eris and Makemake, we find that different types of methane can have the same D/H ratio. Primordial and thermogenic methane can overlap, as can thermogenic and abiotic methane (Figure 2). This makes it difficult to use the D/H ratio alone to uniquely resolve the origin of methane on Eris and Makemake, unless the D/H ratio happens to be very high or very low. Overall, we predict the following D/H ratios for methane on these worlds: primordial = $(5.6-30.1)\times10^{-4}$, thermogenic = $(1.1-12.3)\times10^{-4}$, and abiotic = $(0.8-4.9)\times10^{-4}$. These ranges follow from the expectation that Eris and Makemake water will have D/H ratios within the JFC range given above.

**Figure 2.** D/H ratios for different types of methane that may represent methane on the surfaces of Eris and Makemake. Primordial methane would be accreted in the building blocks of these bodies, or delivered by comets throughout their histories. Thermogenic methane would be produced from accreted organic matter that was "cooked" in the rocky cores of these bodies. Abiotic methane would be a product of $H_2$ reacting with $CO_2$ or CO in hydrothermal systems at the base of a subsurface ocean. In our model, the D/H ratio of methane depends on the D/H ratio of water (see Section 2). Our current best estimate for the D/H ratio of water on Eris and Makemake is within the range encompassed by Jupiter-family comets (JFCs). The total 1σ range for Eris/Makemake is also shown (see Figure 3 for individual comparisons).

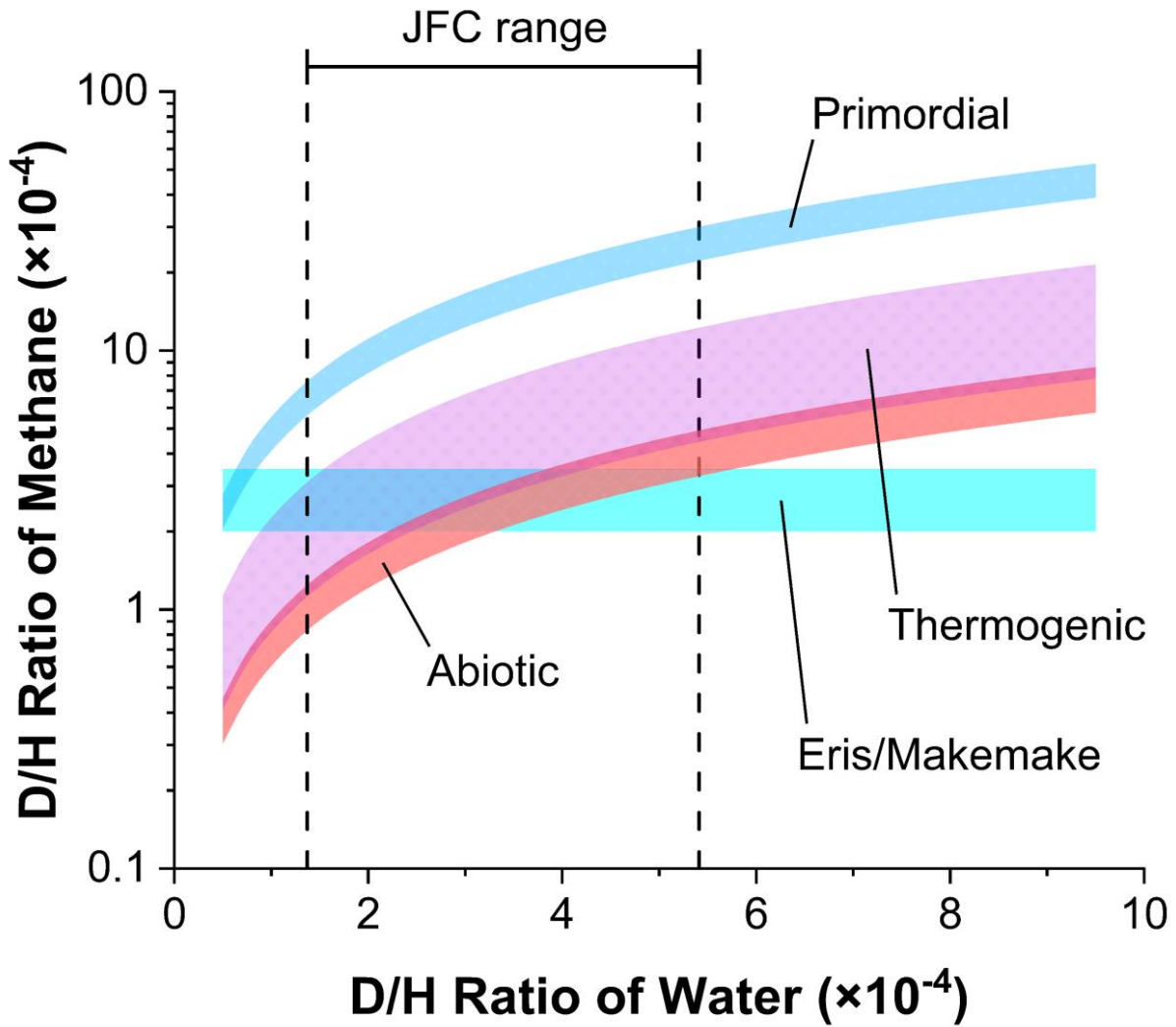

We can now test which types of methane have D/H ratios that are compatible with values derived by Grundy et al. (2024b) for Eris and Makemake. Figure 3 shows how our model predictions compare to the observational data. It is evident that D/H ratios for primordial methane are too large. The fractionation factor for primordial methane in the outer solar nebula would need to be significantly smaller than what we currently think it is (see Section 3.3). Alternatively, isotopically light water would be implied with a D/H ratio less than $8.5×10^{-5}$ (Figure 2), unlike known comets (see Biver et al., in press). These discrepancies suggest (but do not prove) that the present surface inventories of methane on these TNOs are not remnants of an accreted inventory of methane. In other words, Eris's and Makemake's methane do not appear to have a primordial origin. Instead, we find that the moderate observed values of the D/H ratio are consistent with their methane being abiotic or thermogenic. The data fall in the middle of the abiotic range and at the lower end of the thermogenic range (Figure 3).

**Figure 3.** Comparisons between predictions (Figure 2) and observations (Grundy et al., 2024b) of the D/H ratio of methane on the surfaces of Eris and Makemake. Ranges are 1σ.

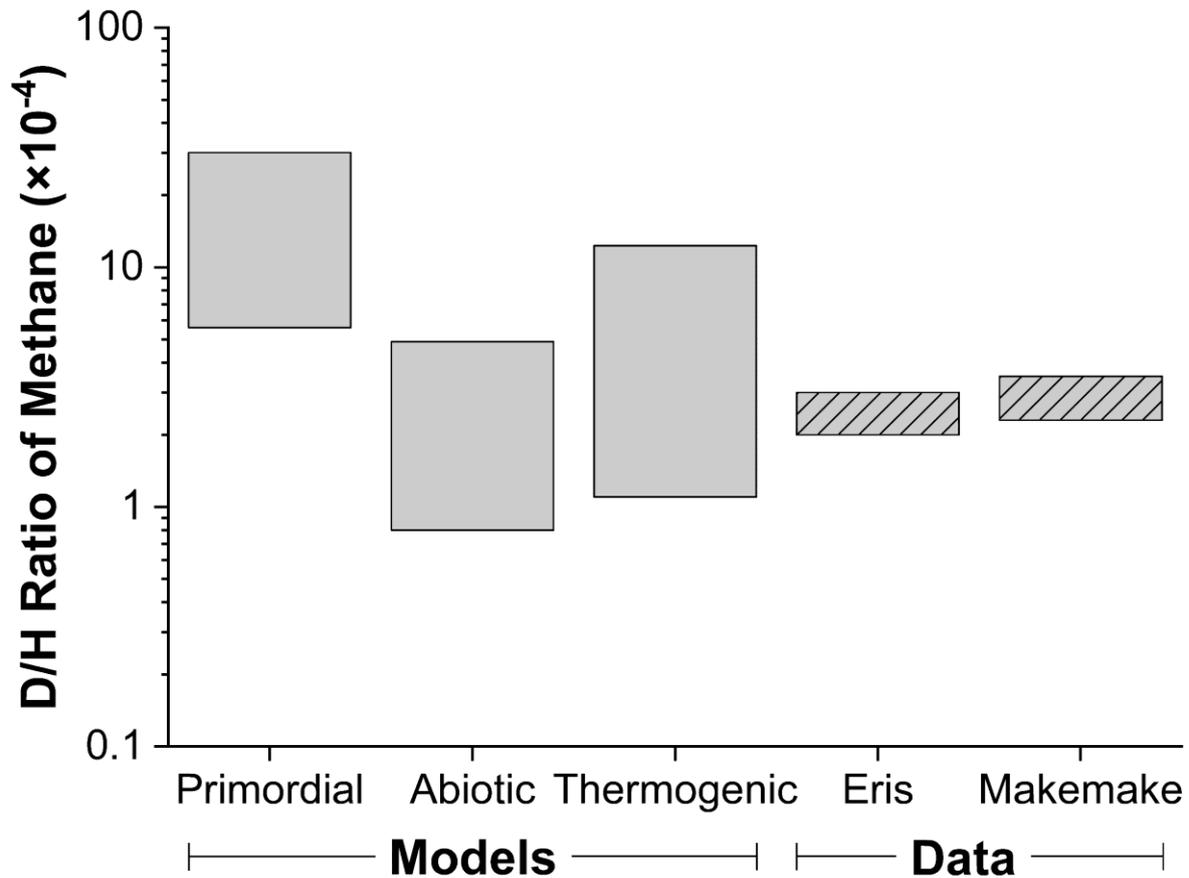

Because Eris and Makemake are both large TNOs with indistinguishable D/H ratios, it is tempting to assume that they have the same type of methane source. This does not have to be true, but it is our working hypothesis until there are data that show otherwise. It should also be noted that, while we have focused on single sources of methane for simplicity in discussing hypotheses, we cannot rule out more complex scenarios involving mixed sources on Eris and Makemake (e.g., Glein, 2023), including mixtures with primordial methane. As an example, a makemakean mixture could contain as much as ~56% of primordial methane, if it were mixed with the lowest D/H abiotic methane from our model. Conversely, undetectable (<10 ppm) levels of CO on the surfaces of these bodies may limit the quantity of primordial $CH_4$ that could be mixed in, given that CO is usually significantly more abundant than $CH_4$ in cometary comae (Dello Russo et al., 2016; Lippi et al., 2021). The exception would be if CO were destroyed through aqueous reactions in the interior, or if it is buried under more volatile ices such as $N_2$ (Glein and Waite, 2018). The hypothesized CO removal mechanism would need to be stronger on Eris/Makemake than on Pluto, where some CO ice is seen (Owen et al., 1993; Grundy et al., 2016).

We chose not to give a solar methane endmember the same prominence as other types of methane owing to a lack of observational evidence. "Solar" methane means that there could be accreted methane with a protosolar D/H ratio ($2.2×10^{-5}$; Geiss and Gloeckler, 2003; Aléon et al., 2022). No comets are known to have such a low D/H ratio (see above), although it has been proposed that their D/H ratios may be indicative of mixing between interstellar and solar sources (Cleeves et al., 2014; Alexander et al., 2017a). We build upon this idea in the Appendix and discuss its implications in Section 3.3. It may be that comets do not have near-solar D/H ratios because they formed far enough away from the Sun, which made the interstellar contribution large. Regardless of the reason, even if nearly pure solar methane had been present in some regions of the outer solar nebula, its D/H ratio would be much less than and thus inconsistent with observed values ($(2.0-3.5) \times 10^{-4}$) on Eris and Makemake (Grundy et al., 2024b).

*3.2. D/H as a window into the subsurface*

What does the inferred presence of abiotic or thermogenic methane mean for conditions inside Eris and Makemake? Here, we focus on temperature since it is a critical link between the geophysical and geochemical evolution of icy worlds. To form abiotic methane in hydrothermal systems, an interior ocean of liquid water can be presumed to promote extensive water-rock interaction at high enough temperatures, which are needed to overcome kinetic barriers to abiotic synthesis (McCollom and Seewald, 2007). However, temperatures cannot be too high; otherwise, methane would not be thermodynamically stable and thus not produced (Shock, 1992). A synthesis of existing laboratory and thermodynamic constraints suggests that a reasonable range for the abiotic formation of $CH_4$ is ~200-400°C (Wang et al., 2018). The lower limit is not sharp as it depends on the residence time of hydrothermal fluids in heated regions below the seafloor. The residence time of putative hydrothermal fluids on Eris and Makemake could be longer than on Earth because of slower fluid flow due to lower gravity (e.g., Glein et al., 2008). Depths of hydrothermal circulation on Eris may be expected to be similar to those on Earth, while deeper circulation could occur on Makemake (Vance et al., 2007). Thus, the length scale will either cancel out or proportionally increase the relative residence time. We should also be aware that the widely held notion of low-temperature (<90°C) abiotic methane from the Lost City hydrothermal system has been challenged by clumped-isotope geothermometry, which suggests that Lost City's abiotic methane actually formed at ~200-370°C (Wang et al., 2018).

As an additional consistency check, we can evaluate whether observed D/H ratios on Eris and Makemake are compatible with isotopic equilibrium between methane and water (e.g., Glein et al., 2009) over the proposed temperature range. This is a more restrictive case for abiotic methane than that considered in Table 1. The equilibrium fractionation factor ($\alpha_{CH4-H2O}^{abio}$) is between 0.87 (at 200°C) and 0.90 (at 400°C) (Horibe and Craig, 1995; Turner et al., 2021). Repeating the analysis that we performed in Section 3.1, we deduce that abiotic methane at isotopic equilibrium with hydrothermal fluids would have D/H = $(1.2-4.9) \times 10^{-4}$, consistent with the 1σ range of $(2.0-3.5) \times 10^{-4}$ for Eris + Makemake (Grundy et al., 2024b).

Figure 4a illustrates what the context for an abiotic origin of methane might look like. This conceptual cartoon should not be taken too literally. The main message is that an abiotic origin of methane would point to high-temperature "ocean world" activity in the interiors of Eris and Makemake. This type of highly processed and active interior may be conducive to supporting life at some unknown time (e.g., Martin et al., 2008; Waite et al., 2017).

**Figure 4.** Three scenarios for how methane generation might fit into the geophysical evolution of Eris and Makemake. These are radial cross sections. From left to right, the interior becomes cooler and key processes that affect the availability of methane can change. This progression may represent a decrease in size when comparing different bodies, or the evolution of an individual body through time. Subsurface phenomena shown here are hypothetical on Eris and Makemake. Nevertheless, the methane-producing processes are consistent with observed D/H ratios (Figure 3), and other processes may be suggested by drawing parallels between large TNOs and comparable icy satellites (McKinnon et al., 2008). Dimensions not to scale.

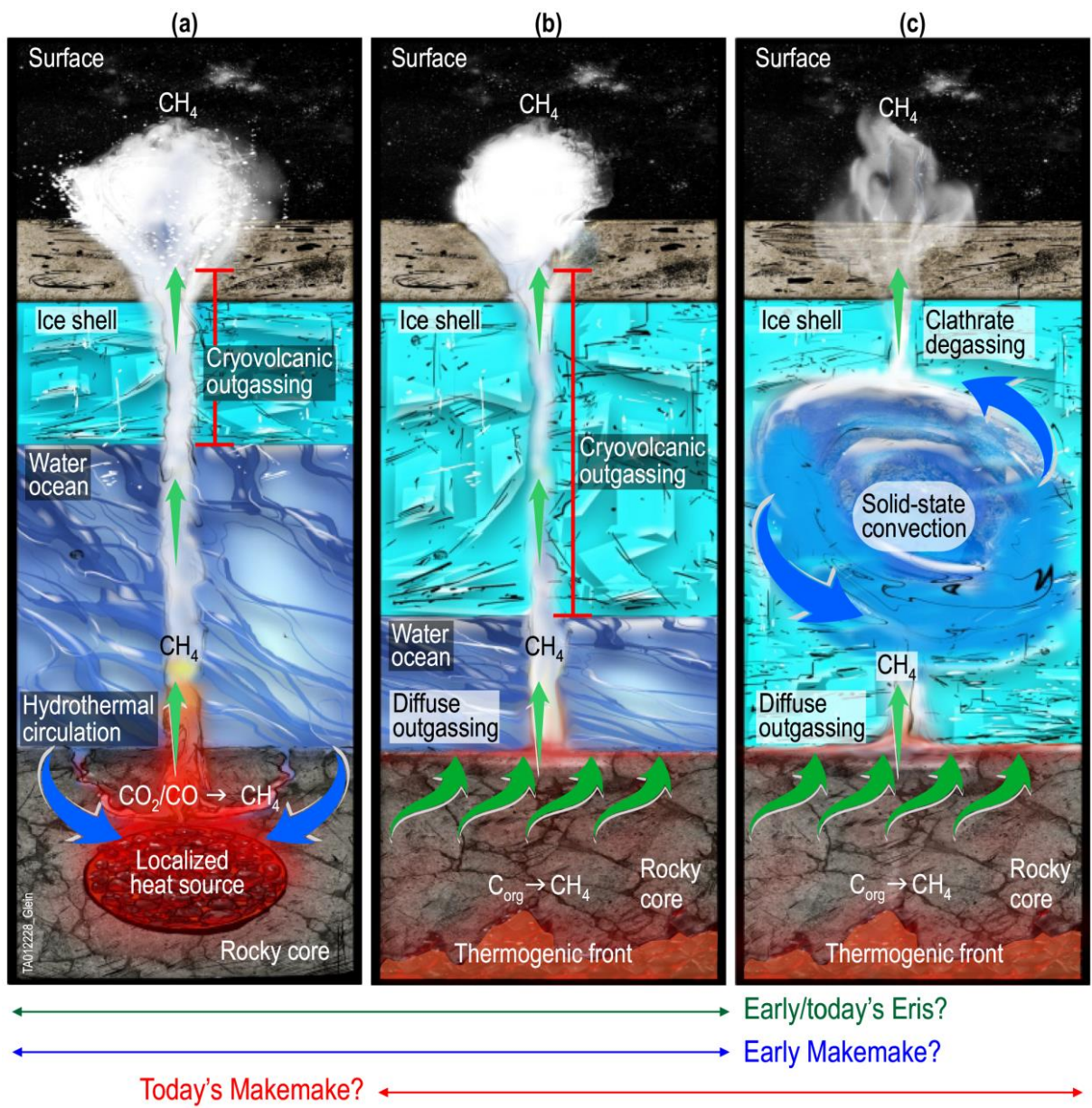

If Eris's and Makemake's methane is thermogenic, then alternative pictures may emerge (Figure 4b, 4c). In these cases, elevated temperatures are still implied, but conditions in the core only need to be warm (>150°C; Stolper et al., 2014) rather than hot. A cooler core would be consistent with a more frozen hydrosphere, as depicted in Figure 4b and 4c. Again, these sketches are meant to be seen in simplified terms to help visualize and put the thermogenic interpretation of D/H ratios into potential geophysical contexts.

Based on the above discussion, it can be concluded that Eris and Makemake should have rocky cores that reached temperatures of at least ~150°C to produce thermogenic or abiotic methane (since the minimum thermogenic temperature is lower than the minimum abiotic synthesis temperature, we adopt the former value as the overall lower limit). Their cores probably have low densities (e.g., ~2500-3000 kg/m$^3$) and may bear some resemblance to mud, as they are expected to be composed of phyllosilicates such as serpentine and may also contain trapped fluids (e.g., Melwani Daswani and Castillo-Rogez, 2022).

The inference that Eris and Makemake have rocky cores that are (or were) warm or hot makes sense in light of other available information. They are both relatively big (~2300 and ~1400 km in diameter, respectively) bodies with bulk densities (>1700 kg/m$^3$; Parker et al., 2018; Holler et al., 2021) that skew toward the high end for icy worlds. They are likely to be rock-rich (Bierson and Nimmo, 2019). This means there will be a relatively large heat supply from the decay of radioactive isotopes in their interiors, which will promote core formation as a consequence of water-rock differentiation (Desch et al., 2009; Loveless et al., 2022) and enable higher temperatures to be reached (core formation may have occurred due to accretional heating as well, at least in the case of Eris). Indeed, it is significant in this respect that Nimmo and Brown (2023) recently found that Eris is likely to have a differentiated interior so that it could become tidally locked to its satellite Dysnomia (Szakáts et al., 2023; Bernstein et al., 2023). This provides an independent line of evidence that Eris's interior could reach temperatures in excess of 0°C, as elaborated below. Additional sources of heat inside Eris and Makemake are serpentinization (Farkas-Takács et al., 2022) and tidal dissipation (Saxena et al., 2018). The large sizes of these bodies also favor heat retention, which implies that higher core temperatures can persist for longer durations.

These expectations can be quantified by thermal evolution modeling. While time-dependent models specific to Eris and Makemake have not been published, we can gain insights into what is plausible on these bodies by considering models that have been developed for Pluto and the uranian satellite Titania as analogues for Eris and Makemake, respectively.

Eris is slightly smaller but considerably denser than Pluto (~2400 kg/m$^3$ vs. ~1850 kg/m$^3$). Assuming a similar composition of rock, Eris's interior should be warmer than Pluto's. Models for Pluto suggest that high core temperatures can be reached, even up to ~1000°C (Kamata et al., 2019; Bierson et al., 2020). Such temperatures, together with the potential for a subsurface ocean (as on Pluto; Nimmo and McKinnon, 2021), would be sufficient to support hydrothermal and metamorphic processes in Eris's core. This information adds further, albeit circumstantial, evidence to the case that Eris's methane is likely to be abiotic or thermogenic. Peak methane production inside Eris may have taken place ~2 Gyr ago when core temperatures could have peaked (e.g., Kamata et al., 2019; Bierson et al., 2020). However, we cannot exclude the possibility of ongoing methane production, as present core temperatures (e.g., ibid) may still be sufficient to produce fluids throughout the methane generation window (~150-400°C; see above). We hypothesize that situations resembling Figure 4a or 4b (or a hybrid) may represent key features of Eris's internal evolution. Grundy et al. (2024b) describe how the presence of $N_2$ on Eris can also be consistent with such scenarios of endogenic volatile production.

Because Makemake is smaller and less rock-rich than Eris, its interior can be expected to be less thermally evolved than Eris's. Nevertheless, Makemake appears to be large enough and to contain enough rock mass to favor water-rock differentiation (Desch et al., 2009; Loveless et al., 2022). Indeed, the threshold for an object in the Kuiper belt to be differentiated seems to be below a radius of ~606 km, the value for Charon – which shows evidence of differentiation and early liquid water activity (Spencer et al., 2021).

We can use results from models for Uranus's moon Titania to more specifically guide our understanding of the thermal evolution of Makemake. Titania (mean radius ~790 km) is slightly larger than Makemake (mean radius ~715 km); their densities also appear to be similar, although Makemake (~1700-2100 kg/m$^3$) may be denser than Titania (~1700 kg/m$^3$). Another reason why Titania makes an attractive point of reference is that models have recently been developed to study its thermal evolution. Those models indicate that core temperatures up to ~600°C could be reached inside a Titania-like body (Bierson and Nimmo, 2022; Castillo-Rogez et al., 2023). The models do not feature tidal heating in a rocky core inside Titania, so they should be relevant to the core evolution of Makemake driven by radiogenic heating. Predicted core temperatures suggest that the situation for hydrothermal activity and organic metamorphism is indeed not as favorable inside Makemake as on Eris, but temperatures could still be sufficient to support the production of abiotic or thermogenic methane. Models also indicate that an interior ocean could form but may not persist until the present-day (ibid). Our interpretation is that the most favorable time to produce abiotic methane on Makemake is early in its history, when the existence of an ocean to support hydrothermal circulation would be more likely. Therefore, the scenario in Figure 4a may represent early Makemake (e.g., ~2-4 Gyr ago) if its methane is abiotic. If Makemake's methane is thermogenic, then it is possible that methane may have been produced at any point (or perhaps continuously at a low level) over a longer period of time, ranging from ~4 Gyr ago (e.g., Figure 4b) to today (e.g., Figure 4c).

The above discussion is premised on models of heat generation by the decay of long-lived radionuclides ($^{40}$K, $^{232}$Th, $^{235}$U, $^{238}$U). Results from those models suggest that temperatures can be high enough to produce abiotic or thermogenic methane in the cores of Eris and Makemake. Thus, there is no need to invoke stronger heating powered by the short-lived radioactivity of $^{26}$Al. Of course, our interpretation of D/H ratios, which suggests production of abiotic or thermogenic methane inside Eris and Makemake, does not preclude a role for $^{26}$Al. The underlying processes (e.g., Figure 4a, 4b) would be more vigorous if such heating occurred. Yet, current geophysical interpretations argue against early accretion of TNOs (<4 Myr after Ca-Al-rich inclusions) to minimize $^{26}$Al-driven heating; otherwise, it would be difficult to understand how Pluto and Charon attained similar bulk densities (McKinnon et al., 2017), and why some Kuiper belt objects have bulk densities near or below that of water ice (Bierson and Nimmo, 2019).

*3.3. Further considerations for primordial methane*

Why don't we see evidence of primordial methane in the D/H ratios of methane on the surfaces of Eris and Makemake? There are several possibilities. First, Eris and Makemake could have formed from building blocks that were deficient in methane, if those building blocks formed at temperatures that were too warm to trap methane in ices (e.g., >75 K; Niemann et al., 2005). In this case, we might assume that Eris and Makemake formed closer to the Sun. Indeed, this would be consistent with dynamical models of how the Kuiper belt was structured, where the ancestral trans-Neptunian disk of planetesimals (prior to giant planet migration) is thought to have been more compact (Morbidelli and Nesvorný, 2020). However, some Jupiter-family comets (the remnant building blocks of Kuiper belt/scattered disk objects) are known to contain methane (e.g., Lippi et al., 2021), which is almost certainly primordial.

An even warmer origin scenario that was proposed for Eris suggests that Eris started to form inward of the "snowline" in the solar nebula to explain how it got a relatively high density (Reynard and Sotin, 2023). If temperatures were marginally adequate to allow water ice condensation (~150 K; Öberg et al., 2011) in Eris's formation environment, then they would have been too high to allow condensation of methane-bearing ices (see above). On the other hand, a more conventional explanation for Eris's high density is ice removal caused by a giant impact (Barr and Schwamb, 2016).

This is consistent with a recent determination of a low bulk density (700±500 kg/m$^3$) for Eris's moon Dysnomia (Brown and Butler, 2023).

Perhaps a more promising possibility is that methane might have been present in the building blocks and existed on Eris and Makemake early in their histories, but was lost to space (or destroyed through hot geochemistry), leaving abiotic or thermogenic processing to make new methane. This scenario is reminiscent of the hypothesis that a secondary source of $N_2$ could have replaced primordial $N_2$ on Triton (Lunine and Nolan, 1992). Primordial methane could have been lost during the accretion process when Eris and Makemake were less massive while they were still growing (Stevenson, 1993). Or, it might have been lost while Eris and Makemake orbited closer to the Sun, when their surface environments would have been warmer and more conducive to escape (e.g., Johnson et al., 2022). Another possibility is that giant impacts could have caused primordial volatiles to be lost (Barr and Schwamb, 2016; Arakawa et al., 2019). Abiotic or thermogenic methane could have been produced later after rocky cores had heated up sufficiently (see Section 3.2). These types of methane could have then been delivered to the surface to replace or at least significantly overprint a primordial methane inventory (Neveu et al., 2015; Menten et al., 2022; Howard et al., 2023). The non-detection of CO on Eris and Makemake by JWST makes escape scenarios appealing (see Grundy et al., 2024b).

Third, our understanding of what the D/H ratio of primordial methane should look like may be incomplete. We can envision how this might happen in two ways. First, it could be argued that primordial methane would exchange hydrogen isotopes with isotopically lighter water in a subsurface ocean, erasing an original heavy D/H signature (Miller et al., 2019). The problem with this scenario is that laboratory experiments show that the exchange reaction is exceedingly slow at likely ocean temperatures. As an example, the mean lifetime of the C-D bond in $CH_3D$ dissolved in water can be calculated to be of order $10^{17}$ yr at 0°C (Turner et al., 2022). This value comes from an Arrhenius extrapolation of higher-temperature (>376°C) rate constants, so it is probably not highly accurate, but the overall magnitude is suggestive that this is not a promising path to pursue unless additional assumptions are introduced.

A different possibility that may be more feasible is that primordial methane might be able to have lower D/H ratios than adopted here (see Figure 2). Perhaps there could be some process that produced less deuterated methane in the solar nebula. There are insufficient data from comets other than 67P to rigorously test this idea. For the most part, only non-diagnostic upper limits on the D/H ratio (<5×10$^{-3}$; Kawakita et al., 2005; Bonev et al., 2009; Gibb et al., 2012) are presently available, although Kawakita and Kobayashi (2009) may have detected $CH_3D$ in comet C/2004 Q2. If they did, then the D/H ratio was (3.8±1.3)×10$^{-3}$, which is similar to the value of (2.41±0.29)×10$^{-3}$ from comet 67P (Müller et al., 2022). HCN in comet Hale-Bopp also has a similar D/H ratio ((2.3±0.4)×10$^{-3}$; Meier et al., 1998). This may be another indication that primordial C-H compounds can be expected to be strongly enriched in deuterium.

What guidance can theory give us? As an example, our mixing model for primordial methane in the outer solar nebula (see the Appendix) suggests that $\alpha_{CH4-H2O}^{OSN}$ may not be substantially smaller than what was found for comet 67P (Figure A1). This more physically-grounded perspective provides justification that the fractionation factor can be close to constant over the parameter space that is relevant to Eris and Makemake. Hence, our general finding that Eris and Makemake would have higher-than-observed D/H ratios if they bear accreted methane appears to be supported (Figure A2), at least for the case where mixing is the dominant process that controls the D/H ratio of methane relative to that of water. We caution that our mixing model relies on assumptions for the D/H ratios of interstellar methane and water, as well as how they might be diluted similarly by materials with protosolar D/H ratios. Values of $\alpha_{CH4-H2O}^{OSN}$ are not yet available from multiple comets to test these assumptions. Therefore, a more conservative interpretation of Figure A1 is that it establishes a plausibility argument – it is reasonable to adopt $\alpha_{CH4-H2O}^{OSN}$ from comet 67P as a broadly relevant parameter. The implication is that we should be careful not to overinterpret predictions from the mixing model.

We can also consider cases where the kinetics of H-D exchange dominates. We focus first on results from models by Mousis et al. (2000; 2002) since they represent the current paradigm for the origin of methane on Titan (e.g., Thelen et al., 2019). Mousis et al. (2002) modeled the D/H evolution of methane in the solar nebula. Their results cannot be directly compared with comet 67P data because they assumed that the initial D/H ratio of methane (($1.1$-$3.2$)$\times10^{-4}$) was much lower than the value for comet 67P. Nevertheless, we can still calculate the depletion factor – the final D/H ratio divided by the initial value. We obtained depletion factors between ~0.4 and ~0.5. These values are not small enough to reproduce Eris's and Makemake's D/H ratios starting from the methane D/H ratio in comet 67P. The depletion factor would need to be ~0.1.

Additional insight can be obtained from the results of Mousis et al. (2000). These researchers studied cometary HCN, but owing to a lack of experimental data for HCN, they used macroscopic rate constants for methane in their solar nebula model. Thus, the behavior of HCN in their model is likely to resemble the behavior of methane. Since Mousis et al. (2000) provided D/H values for both water and HCN, we can study how the fractionation factor between these two species would evolve as they exchange hydrogen isotopes with $H_2$.

Predicted evolutionary paths are shown in Figure 5, where we have substituted $CH_4$ for HCN (see the explanation given above). The D/H ratio of methane can be seen to decrease along the evolutionary pathway, consistent with Mousis et al.'s (2002) model that was dedicated to methane only. It can also be seen that methane evolves much less than water. The reason for this is because water appears to undergo faster H-D exchange with $H_2$ than methane does (Lécluse and Robert, 1994). Since methane can exist as a gas over a wider range of temperatures than water can, the isotopic evolution of methane can be decoupled from that of water. Nevertheless, sluggish exchange kinetics might have inhibited $CH_4$-$H_2$ isotopic equilibrium in the solar nebula at temperatures (e.g., between ~75 and ~150 K; Notesco et al., 1997) where methane was a gas and water was frozen as ice. Thus, strongly deuterated methane could have persisted according to this logic. We might then expect the $CH_4$-$H_2O$ fractionation factor to increase over the evolutionary pathway (see Figure 5), which is in the opposite direction of what is needed to support a primordial origin of methane on Eris and Makemake. For example, we find that if low D/H (Hartley 2-like) water were derived from high D/H (67P-like) water, the D/H of methane would only decrease to a value of ~$1\times10^{-3}$ (Figure 5). This is not enough to achieve consistency with observed values (($2.0$-$3.5$)$\times10^{-4}$) on Eris and Makemake (Grundy et al., 2024b).

**Figure 5.** Predictions for the D/H evolution of methane and water in the solar nebula from a model of coupled chemistry and dynamics (Mousis et al., 2000). It is assumed here that model results for HCN are representative of $CH_4$ (see Section 3.3). Blue lines show model output for different initial (presolar) D/H ratios in methane, and red lines correspond to various values of the D/H fractionation factor between $CH_4$ and $H_2O$ in the outer solar nebula for comparison. Also shown are the D/H values for comet 67P (inside the turquoise box) and the lower end of the 1σ uncertainty range for the D/H ratio of water in comet Hartley 2 (Hartogh et al., 2011). Square data points were extracted from Figure 7 in Mousis et al. (2000).

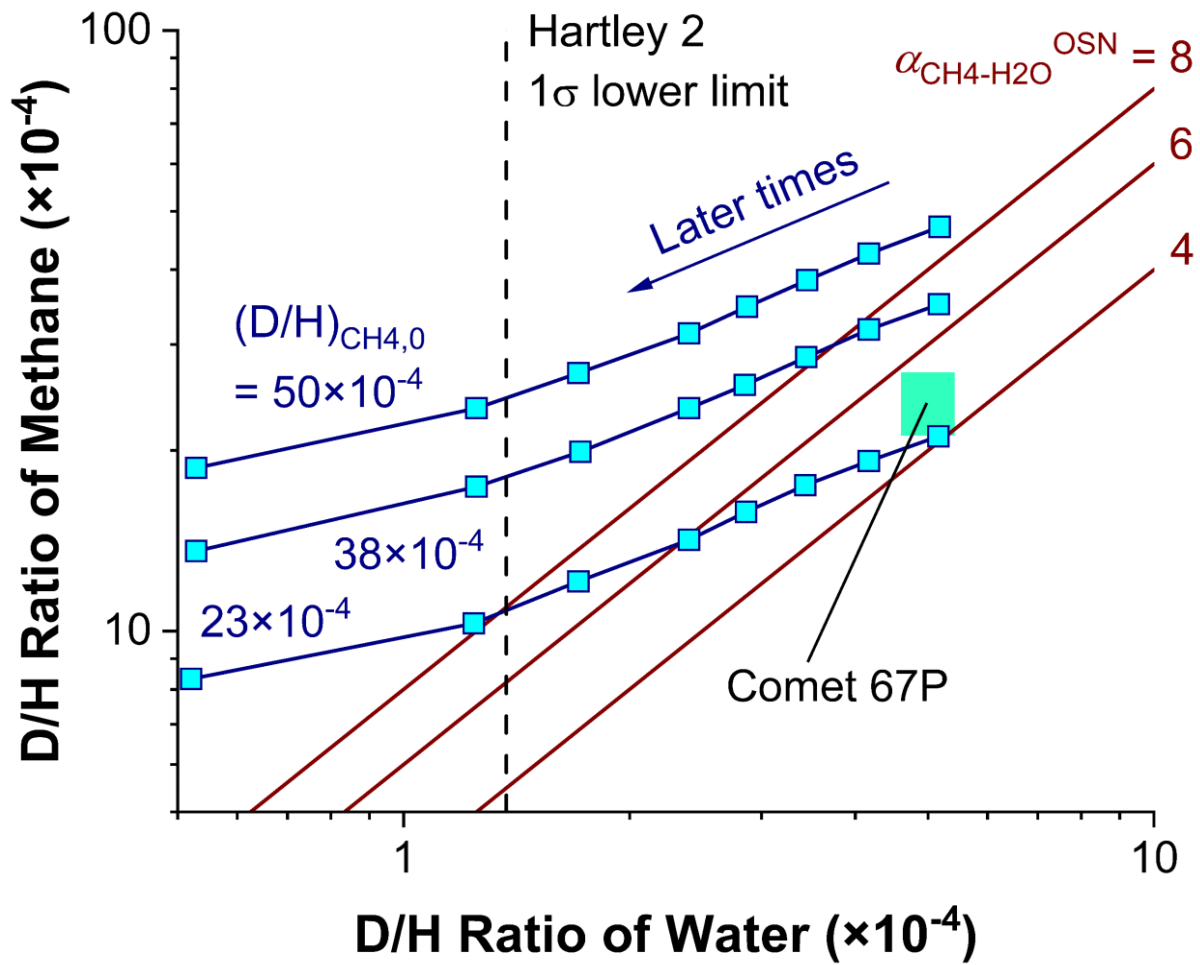

We should mention that Titan's atmospheric methane (D/H = (1.36±0.08)×10$^{-4}$, error-weighted mean from multiple data sets; Nixon et al., 2012) would also plot below the minimum ordinate value in Figure 5, so this model can no longer explain the D/H ratio of Titan's methane. Thus, a primordial origin of methane on Titan may be challenged (see Section 3.5). Because the origin of methane on Titan remains unresolved, Titan's methane cannot be claimed to provide an example of outer solar system primordial methane having a lower D/H ratio than 67P methane.

The isotopic consequences of radiation-driven chemistry should also be considered. Deuterium fractionation in astrochemical environments is an important area of study, aimed at understanding observations of D enrichment in molecular clouds, protoplanetary disks, and primitive materials (Millar et al., 1989; Ceccarelli et al., 2014). Below, we examine results from chemical models that can help us understand alternatives for primordial methane that would be more complicated than our nominal case (see Section 2.2). Several models have been developed (Yung et al., 1988; Aikawa and Herbst, 1999; Willacy and Woods, 2009; Albertsson et al., 2014; Cleeves et al., 2016), but predictions for the D/H ratio of methane are less common. This is presumably because D/H ratios have been determined for other compounds (e.g., $H_2O$, HCN), and the determination for comet 67P methane is recent (Müller et al., 2022).

Yung et al. (1988) were the first to model photochemical D/H fractionation of methane in the solar nebula. From their model of neutral gas chemistry, they computed the D/H ratio of methane as a function of temperature. We found that the D/H ratios for Eris and Makemake can be reproduced at temperatures between ~200 and ~230 K. These temperatures seem too high for where Eris and Makemake are thought to have formed (~20-30 AU; McKinnon et al., 2021). Protosolar disk temperatures at ~20-30 AU have been estimated to range from ~30-80 K (Desch, 2007; Yang et al., 2013; Öberg and Wordsworth, 2019). Also, Eris and Makemake are believed to have significant amounts of ice in their interiors (McKinnon et al., 2008), and water ice condensation requires temperatures below ~150 K (e.g., Öberg et al., 2011). Methane condensation requires temperatures below ~75 K (Notesco et al., 1997) and potentially as low as ~30 K to accrete the pure ice (Schneeberger et al., 2023). Comet 67P data can provide additional context. To reproduce the measured D/H ratio in methane, a temperature of ~135 K would be needed (Yung et al., 1988). This is also higher than what is expected if 67P volatiles were accreted in clathrate hydrates (~50 K; Mousis et al., 2016) or amorphous ice (~70 K; Almayrac et al., 2022). It is possible that mixing of warm and cold methane could give the appearance of cool methane. The simplest example of mixing is considered in the Appendix and found to be consistent with our previous results (Figure A2). Predictions for more complex mixing scenarios will need to be made in the future using chemical-dynamical models of outer solar nebula evolution.

Aikawa and Herbst (1999) also made predictions for the D/H ratio of methane in the outer solar system. Their model builds on the Yung et al. (1988) model by including reactions with ions and accounting for ice formation. It was found that the D/H ratio should have a value of ~$3\times10^{-2}$ and not vary appreciably between 30 and 123 AU from the central star (Aikawa and Herbst, 1999). The predicted D/H ratio is too high to be consistent with values for Eris and Makemake (($2-3.5)\times10^{-4}$; Grundy et al., 2024b). The predicted value is also higher than the measured value for comet 67P (($2.41\pm0.29)\times10^{-3}$; Müller et al., 2022). This is a problem as comet 67P methane is a known case of primordial methane. The Aikawa and Herbst (1999) model requires revision before it can provide an accurate description for the D/H ratio of primordial methane. It remains to be seen whether modifications could make the model consistent with comet 67P data and allow the data from Eris and Makemake to be explained.

The most recent model of primordial methane may be most relevant. Cleeves et al. (2016) accounted for thousands of reactions and grain-surface chemistry in their protoplanetary disk model. Like Aikawa and Herbst (1999), Cleeves et al. (2016) found that methane is more prone to deuteration than water is. This supports our assumption of a large value for the $CH_4$-$H_2O$ fractionation factor ($\alpha_{CH4-H2O}^{OSN}$; see Table 1). Although we have suggested that the D/H ratio of methane might reflect an interstellar contribution (see the Appendix), Cleeves et al. (2016) showed that ion-molecule reactions in the outer solar nebula can cause methane to acquire large D enrichments. Their standard model predicts that the D/H ratio of methane can reach ~$2\times10^{-3}$ in the ~20-40 AU region. This is intriguingly similar to comet 67P ($(D/H)_{CH4}$ = $(2.41\pm0.29)\times10^{-3}$; Müller et al., 2022). If this model represents reality, then primordial methane should be strongly deuterated (D/H > $1\times10^{-3}$) throughout the outer solar system (Cleeves et al., 2016). The D/H ratio of primordial methane would be too high to be consistent with that on Eris/Makemake (($2-3.5)\times10^{-4}$; Grundy et al., 2024b). On the other hand, Cleeves et al. (2016) found that shorter times of exposure (on the order of $10^5$ yr rather than $10^6$ yr) to ionizing radiation would lead to less buildup of deuterium in methane. In this case, primordial methane can have lower D/H ratios that overlap with those observed on Eris and Makemake. However, it is unlikely that prolonged radiation exposure can be avoided given the long formation times (>4 Myr) of Kuiper belt objects (McKinnon et al., 2017; Bierson and Nimmo, 2019).

Overall, we find that the presence of primordial methane is disfavored on Eris and Makemake. The only incontrovertible example of primordial methane in comet 67P has a much higher D/H ratio than Eris/Makemake methane, and we expect this to be true for primordial methane in general. Nonetheless, the solar nebula was a complex environment and there may be scenarios that allow primordial methane to acquire D/H ratios that would be consistent with those measured in Eris/Makemake methane, although we have not identified any at this time. What we do know favors scenarios of endogenic methane production (see Sections 3.1 and 3.2). Such non-primordial origins of present-day volatiles also provide the simplest explanation for the absence of detectable CO on Eris and Makemake (see Grundy et al., 2024b). The clearest way to test for a primordial origin of volatiles would be to search for argon and krypton on Eris and Makemake. We predict that they will be like Titan in terms of being impoverished in non-radiogenic noble gases (Niemann et al., 2010; see Section 3.5).

*3.4. Possible effects of unseen surface/atmospheric processes*

A complication could arise if $CH_3D$ and $CH_4$ are fractionated on the surfaces of Eris and Makemake, as a result of seasonal atmospheric cycling (Hofgartner et al., 2019; Young et al., 2020). In this case, the observed D/H ratio would correspond to the uppermost surface, and the D/H ratio of bulk surface methane could differ from the measured value. It is difficult to find support for this kind of effect for several reasons. First, we do not know the morphology of methane deposits on these bodies and whether they would be reflective of layered deposition that might support fractional crystallization (e.g., Glein and Waite, 2018), or if convective mixing (à la Sputnik Planitia on Pluto; McKinnon et al., 2016) might be acting to homogenize the composition. Second, we are not aware of any measurements of isotope fractionation factors between methane gas and solid methane at relevant temperatures (i.e., ~30-40 K; Sicardy et al., 2011; Ortiz et al., 2012). Third, it would be quite a coincidence if Eris's and Makemake's methane inventories look abiotic or thermogenic on the basis of their D/H ratios (Figure 3), but are primordial methane that happens to be fractionated just right on both bodies. This cannot be discounted but seems unnecessarily complex. Fourth, at least on Eris, we tend to disfavor volatility as the main factor driving the observed composition because abundant $CH_4$ is observed (see Section 1) despite $CH_4$ being orders of magnitude less volatile than $N_2$ (Grundy et al., 2024a). If the stratigraphy of surface ices were ordered by vapor pressure, then we might expect $N_2$ to be at the top and to potentially obscure $CH_4$ in deeper layers. This is evidently not the case (Grundy et al., 2024b).

The last issue that needs to be discussed is the potential for secular D/H evolution of methane at the surface. In the Titan literature (from which we draw guidance), a great deal of effort has gone into constraining how much the D/H ratio could have changed from its original value (Pinto et al., 1986; Lunine et al., 1999; Cordier et al., 2008; Mandt et al., 2009, 2012; Nixon et al., 2012). The most important general finding that can be applied to our analysis is that the D/H ratio increases through time. On Eris and Makemake, both photochemistry (in a perihelion atmosphere and presumably in surface ices) and escape should preferentially remove $CH_4$ relative to $CH_3D$, which would enrich the remaining inventory in deuterium.

We do not know how long the observed methane inventories have existed on the surfaces of Eris and Makemake. If volatility effects are of secondary importance (see above), then the D/H ratios of present surface methane would correspond to upper limits for methane that was originally emplaced on the surface. Primordial methane would be even more discrepant, whereas both abiotic and thermogenic methane can still be consistent with the data (Figure 3). If D/H evolution has been substantial, one might wonder whether thermogenic methane may be less viable. This would be of greater relevance to Makemake, whose closer heliocentric distance and smaller mass promote methane photochemistry and particularly escape relative to that at Pluto (Schaller and Brown, 2007; Brown et al., 2011; 2015).

Grundy et al. (2024b) determined $^{12}C/^{13}C$ ratios in methane ice on Eris (~71-100, 1σ) and Makemake (~77-143, 1σ) using JWST data, and found that they are approximately "Earth-like" and consistent with expected values if their methane inventories were derived from $CO_2/CO$ or solid organic carbon. The latter consistencies are significant as they suggest that the surfaces of these bodies do not show a large enrichment of $^{13}CH_4$ due to prodigious escape of $^{12}CH_4$. Because $^{13}CH_4$ and $CH_3D$ have almost identical masses, the apparently unfractionated $^{13}C/^{12}C$ (heavy/light) ratios may imply that the D/H ratios are also not strongly enriched in the heavy isotope (e.g., Mandt et al., 2009; 2012), if escape has been the dominant loss process for methane. This does not mean that escape has not occurred for a long period of time (see Grundy et al., 2024b for further discussion), but it suggests that the overall isotopic consequences of escape are not anomalously large. Since the $^{13}C/^{12}C$ ratio has not been enriched by ~50% or more, then the current D/H ratio is probably also not enriched by a similarly large amount. This may give us confidence that observed D/H ratios can be used to make inferences about the origin of methane (escape is unlikely to have overprinted the original D/H signature). However, it is possible that methane loss from these worlds has been dominated by radiation chemistry rather than escape. Photochemical reactions exert a stronger isotope effect on hydrogen than carbon isotopes (e.g., Nixon et al., 2012), so the D/H ratio could be more enriched than the $^{13}C/^{12}C$ ratio. In this case, current D/H ratios may serve as upper limits for sources of endogenic methane.

*3.5. Relationship to Titan*

Our suggestion that methane on Eris and Makemake is abiotic or thermogenic invites a comparison to Titan, where the origin of methane has been a long-standing mystery (e.g., Owen, 1982). It has been proposed that Titan's methane could be primordial (Mousis et al., 2009a), abiotic (Glein, 2015), or thermogenic (Miller et al., 2019). The origin of methane on Titan has not been revisited since much more detailed cometary data became available thanks to the *Rosetta* mission. Those data appear to disfavor a primordial origin of methane on Titan for two reasons – one based on the D/H ratio and the other based on the $^{84}Kr/CH_4$ ratio.

Titan's atmospheric methane has a relatively low D/H ratio of $(1.36\pm0.08)\times10^{-4}$ (Nixon et al., 2012). We can apply the same analysis to Titan as we did for Eris and Makemake. For the case of Titan, it can be assumed that the D/H ratio of water is within the range of measurements from Enceladus's plume (Waite et al., 2009) and water ice observed in the Saturnian system (Clark et al., 2019). This range is $(1.5-4.4)\times10^{-4}$. The D/H ratio of water ice on Titan itself has not been measured. We use data from elsewhere in the Saturnian system because they are likely to be relevant to Titan. For the adopted range of $(D/H)_{H_2O}$ for Titan, our model (see Section 2) makes the following predictions for the D/H ratio of methane: $(6.1-24.4)\times10^{-4}$ (primordial), $(1.2-10.0)\times10^{-4}$ (thermogenic), and $(0.9-4.0)\times10^{-4}$ (abiotic).

We find that predicted D/H values for primordial methane are significantly higher than the observed D/H ratio on Titan (Figure 6). We interpret this discrepancy as evidence against a primordial origin of Titan's methane. Like the cases of Eris and Makemake, a pre-accretionary process that is currently not indicated by cometary observations would need to be invoked to explain how primordial methane could attain a low enough D/H ratio (see Section 3.3). We note that primordial methane according to our mixing model (see the Appendix) could have D/H down to $5.4\times10^{-4}$ for Titan, but this is still well short of what is needed to explain Titan data. We also note that some amount of deuterium enrichment due to atmospheric evolution on Titan (Pinto et al., 1986) would make the discrepancy larger, as the source of Titan's methane (prior to photochemistry and escape) would then need to have a D/H ratio that is lower than the present Titan value. The maximum modeled extent of atmospheric evolution of the D/H ratio on Titan is represented by the bottom of the Titan bar in Figure 6. Because the extent of atmospheric evolution is unknown, the full Titan range, which models must overlap, extends from the observed D/H ratio down to the lower limit defined by Mandt et al. (2012).

**Figure 6.** Comparisons between Titan model predictions and the inferred range for the D/H ratio of originally outgassed methane on Titan. The latter range is defined by the 1σ upper limit for present-day atmospheric methane (Nixon et al., 2012), and its lower bound was estimated using a model of atmospheric evolution constrained by the $^{12}C/^{13}C$ ratio of Titan's methane (Mandt et al., 2012). The 1σ range for surface methane on Eris and Makemake as a pair (Grundy et al., 2024b) is shown for comparison to the Titan data.

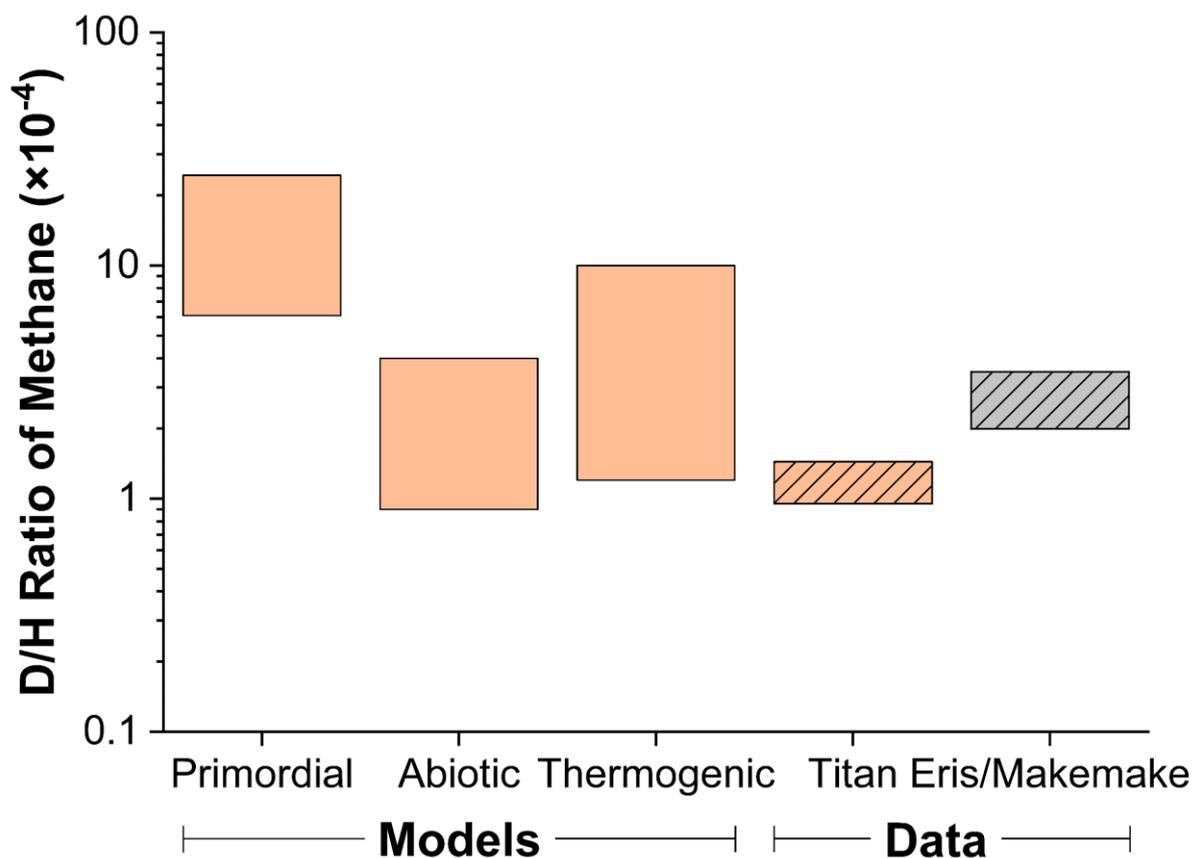

Moreover, Titan's atmosphere is poor in krypton; the upper limit on the ratio of the most abundant isotope, $^{84}Kr$, to $CH_4$, is $7\times10^{-7}$ (Niemann et al., 2010). This is a key clue to the origin of methane as the similar behaviors of Kr and $CH_4$ (in terms of volatility) makes them difficult to fractionate on or inside Titan (Glein, 2015). Kr would be accreted if $CH_4$ was (Mousis et al., 2009b), and the present $^{84}Kr/CH_4$ ratio should be similar to or higher than the primordial value (it could be higher because photochemical depletion of $CH_4$ increases the $^{84}Kr/CH_4$ ratio in Titan's atmosphere).

A variety of processes have been proposed to sequester heavy noble gases on Titan (see Thomas et al., 2007; Jacovi and Bar-Nun, 2008; Mousis et al., 2011; Tobie et al., 2012; Hodyss et al., 2013) and these processes are compatible with conditions on Titan, but the $^{84}$Kr/CH$_4$ ratio can be expected to be a robust tracer since the effects on $^{84}$Kr and CH$_4$ are likely to cancel out. Previously, we did not know what the primordial ratio might be. After *Rosetta* discovered cometary Kr (Rubin et al., 2018), we now have a primordial value ($^{84}$Kr/CH$_4$ = (8.4±4.1)×10$^{-5}$; Rubin et al., 2018; 2019) that can be compared to the Titan upper limit. Titan's atmosphere is depleted in $^{84}$Kr/CH$_4$ by a factor of at least ~60 versus what we would expect to find if methane was accreted. This assumes, of course, that the comet 67P value for the $^{84}$Kr/CH$_4$ ratio is representative of ices containing primordial methane, which cannot be verified as of yet.

Each of the above arguments is not decisive because key data are still lacking, but we consider the combination to be suggestive that Titan's methane is not primordial. Endogenic processes (i.e., abiotic/thermogenic synthesis) that have been argued to be plausible (Glein, 2015; Miller et al., 2019) appear to be implicated instead (see Figure 6). If this interpretation is correct, Titan, Eris, and Makemake may share a kinship in how they acquired methane. At present, we can only speculate on the reason why Eris/Makemake methane (~2.7×10$^{-4}$) has higher D/H ratios than Titan's (~1.4×10$^{-4}$). It may be that similar methane generation processes operated on them, but Eris/Makemake accreted water and/or organic matter with higher D/H ratios, or greater quantities of organic matter, as a result of forming farther from the Sun.

## 4. Concluding remarks

We constructed models of deuterium fractionation to interpret data on D/H ratios measured by the *James Webb Space Telescope* (Grundy et al., 2024b). Our goal was to narrow down the possibilities for the origin of methane on dwarf planets Eris and Makemake. We needed to develop a new geochemical framework for the D/H ratios of primordial, abiotic, and thermogenic methane on icy worlds (see Section 2). To the best of our knowledge, ours is the most comprehensive framework that has been attempted. Our framework builds upon pioneering studies on Titan, and leverages as much empirical data as possible (from Earth, carbonaceous chondrites, and comets) to derive current best estimates for D/H ratios that we would expect to observe if methane is primordial, abiotic, or thermogenic (Figure 2). It should be noted, however, that a limitation of relying on observations of analogues (rather than performing detailed modeling of primordial, abiotic, and thermogenic methane) is that our model predictions probably do not cover all possibilities. This calls for follow-up studies (see below).

After examining a parameter space based on relevant observations, we found that abiotic and thermogenic methane are consistent with Eris's and Makemake's D/H ratios (see Section 3.1). Primordial methane (as currently understood) is not; it would have D/H ratios that are too high. To produce abiotic or thermogenic methane, interior temperatures need to be warm or hot (~150-400°C). This means that a rocky core (likely hydrated and low in density; e.g., ~2500-3000 kg/m$^3$) should be present to allow sufficient radiogenic heat to build up. This inference is consistent with available thermal evolution models, which indicate that rock-rich bodies similar in size to Eris and Makemake can differentiate and experience sufficient heating to facilitate hydrothermal and metamorphic processing of CO$_2$/CO and organic carbon, respectively (see Figure 4). Our inference is also consistent with the geophysical evidence of Nimmo and Brown (2023).

The apparent insignificance of primordial methane on Eris and Makemake seems to suggest that accreted methane was lost to space early in the histories of these bodies, or was perhaps destroyed in an early hot environment. We tried to assess using available literature whether primordial methane might actually have low enough D/H ratios but could not come up with a convincing scenario. We also considered the role of methane loss from Eris's and Makemake's most recently outgassed inventories, which may mean that observed D/H ratios should be seen as upper limits. Yet, this uncertainty may not affect our interpretation unless the amount of D/H evolution was large, and such evolution seems inconsistent with the lack of a large enrichment in the $^{13}C/^{12}C$ ratio of methane ice on Eris and Makemake (Grundy et al., 2024b). Another uncertainty is that it is not clear at this time how representative measured isotopic ratios are of bulk values if surface processes might be redistributing volatiles.

It must be acknowledged that, while our findings provide a clue that methane can be produced inside TNOs and delivered to their surfaces, this does not necessarily mean that all methane on TNOs was internally produced. Each world may have its own story. The general issue of how and when TNOs gain and lose methane is an outstanding question that is central to the origin and evolution of these bodies, and fully addressing it will require more comprehensive investigations. On the other hand, the D/H ratio and non-detection of krypton in Titan's atmosphere seem to point in a direction away from primordial methane being present on that body as well. This may be a hint, as yet speculative, of a common origin of methane on large icy worlds.

We recommend a series of activities to improve the interpretation of D/H ratios of methane on Eris and Makemake. To enable cross comparisons of the origin of methane on large TNOs, JWST should try to measure D/H and $^{13}C/^{12}C$ ratios of methane on Pluto (and Triton). We also need JWST measurements of the D/H ratio of water ice on TNOs to better understand this model input (see Figure 2). Bright TNOs with strong water absorptions (Charon, Haumea, Orcus; Brown et al., 2012) would make excellent targets. Measuring the D/H ratio of water ice on Titan's surface must await *Dragonfly* (Barnes et al., 2021). As discussed in Section 3.3, we do not know how variable the D/H ratio of cometary methane might be. To discern a pattern, it is necessary to determine the D/H ratio of methane in more than one comet. Of particular interest are comets that are known to have a low D/H ratio in water (e.g., comet 45P; Lis et al., 2013). It would be valuable to compare their $CH_4/H_2O$ fractionation factors to that of comet 67P, a high D/H comet. Measurements could be made on a bright comet using ground-based spectroscopy or JWST, or an *in situ* spacecraft mission equipped with a high-resolution (>5800 $m/\Delta m$) mass spectrometer (e.g., Waite et al., submitted) could be sent to a suitable comet. Primordial methane can also be better understood by making predictions of D/H fractionation between $CH_4$ and other volatiles using higher fidelity models of the solar nebula. Such models could be informed by JWST protoplanetary disk observations and validated against comet 67P data (i.e., Müller et al., 2022).

We also lack clarity on key characteristics of cometary organic matter (see Table 1). The D/H ratio is an obvious target, and the organic content of bulk cometary matter is unknown. Refractory grains released from comets Halley and 67P suggest large organic enrichments (Zolotov, 2020). However, CI chondrites contain lower abundances of organic matter (by a factor of ~8), and it is increasingly being recognized that they probably originated farther out in the solar system than previously thought (Gounelle et al., 2006; Desch et al., 2018; Nakamura et al., 2023). The question is whether there was a large organic gradient with heliocentric distance or time in the protoplanetary disk, or if our current data from comets may not be representative of bulk material. A comet surface sample return mission could resolve this dilemma. Detailed characterization of cometary organic matter would also permit comparisons to chondritic organic matter, offering new insights into the relative roles of interstellar/protoplanetary disk processes and aqueous alteration that can occur after the formation of parent bodies.

Closer to home, the next steps are to perform geophysical and geochemical modeling that are tailored to Eris and Makemake. This will allow us to better understand the thermal evolution of these worlds, and quantify how differentiation, metamorphism, hydrothermal circulation, cryovolcanism, sublimation-deposition cycles, escape, and other processes could have shaped the volatile inventories that we see today. Coupled systems modeling would provide more robust forward-model results for interpreting the observed composition. These efforts would set the stage for laboratory experiments under plausible eridian and makemakean conditions to test our models of D/H fractionation, identify limitations in relying on analogies as done in the present paper, and constrain the kinetics of methane production.

In the longer-term, visits to Eris and Makemake with spacecraft are intriguing to ponder. While these worlds are currently far away (Makemake: ~53 AU, Eris: ~95 AU), their potential for endogenic activity (e.g., Figure 4) that appears to show compositional expression on their surfaces is too tantalizing to ignore (Grundy et al., 2024b). JWST has given us a hint, though it must be viewed with caution until we can see the geologic context as demonstrated at Pluto (Moore and McKinnon, 2021). Zangari et al. (2019) studied spacecraft trajectory options and found that Makemake could be reached in ~14 years, which is comparable to the cruise duration of *New Horizons* to Arrokoth (13 yr). A flyby mission to Eris would take ~21 yr. Next-generation launch vehicles together with advanced upper stages may enable these cruise times to be shortened by a few years (nuclear propulsion would be a game changer but may not be available for planetary missions in the foreseeable future). Multi-decade durations are certainly challenges for all but the most junior authors here (see Brandt et al., 2022 for another example), but there is scientific motivation to explore this next frontier, which may be more internally evolved (Kareta, 2023) and perhaps habitable than suggested by previous models of primordial volatile retention (e.g., Schaller and Brown, 2007; Glein and Waite, 2018).

**Acknowledgments**

This work is based in part on observations made with the NASA/ESA/CSA *James Webb Space Telescope*. The data were obtained from the Mikulski Archive for Space Telescopes at the Space Telescope Science Institute, which is operated by the Association of Universities for Research in Astronomy, Inc., under NASA contract NAS 5-03127 for JWST. These observations are associated with JWST Cycle 1 GTO programs #1191 and #1254, and would not have been possible without the incredible efforts that went into making this mission such a success. We thank Heidi Hammel for providing observing time for the Makemake program. We gratefully acknowledge two reviewers for providing constructive feedback. C.R.G. was supported by NASA grant NNN13D485T (Habitability of Hydrocarbon Worlds: Titan and Beyond) and SwRI IR&D grant 15-R6321. J.I.L. acknowledges support from NASA grant NNX17AL71A. Special thanks to Gabriel Rodriguez and A.J. Galaviz for their help and patience in preparing figures. C.R.G. also thanks Conel Alexander, George Cody, and Dionysis Foustoukos for helpful discussions on chondritic insoluble organic matter, Ben Teolis and Ujjwal Raut for clarifying how methane ice could be subject to chemical and escape D/H fractionation, and Danna Qasim and Rosario Brunetto for providing some references. C.R.G. dedicates this paper to the memory of C.G., who was a joy and is missed.

**Appendix. A possible physical counterpart of the empirical fractionation factor: Outer solar nebula methane as a mixture of interstellar and solar methane**

Following Alexander et al.'s (2017a) treatment of cometary water as a mixture comprised of interstellar and solar sources, here we show how the isotopic composition of cometary methane can be treated in an analogous manner. "Solar" means methane or water equilibrated with $H_2$ in warm regions of the solar nebula (e.g., Yang et al., 2013), and "interstellar" is a D-rich endmember reflecting cold conditions in the presolar molecular cloud (e.g., Qasim et al., 2020).

For typical environments in which H is much more abundant than D, the D/H ratio (denoted by $R$ to avoid clutter) of the mixture that would be observed is given by the contributions from interstellar (int) and solar (sol) endmembers

$$R_{obs} = f_{int}R_{int} + f_{sol}R_{sol} = f_{int}R_{int} + (1-f_{int})R_{sol},$$ (A1)

where $f_i$ is defined as the fraction of the fully protiated isotopologue contributed by the $i$th source. If we combine isotopic mass balance equations for methane and water to eliminate the fractions as explicit variables, we can derive an equation for the fractionation factor between methane and water in the outer solar nebula. This is shown below

$$\alpha_{CH_4-H_2O}^{OSN} = \frac{R_{CH_4,obs}}{R_{H_2O,obs}} = \frac{R_{sol}}{R_{H_2O,obs}} + \frac{(R_{CH_4,int} - R_{sol})}{(R_{H_2O,int} - R_{sol})} \times \frac{(R_{H_2O,obs} - R_{sol})}{R_{H_2O,obs}}.$$ (A2)

It should be clarified that this equation is valid only in the limit where methane and water have the same $f_{int}$ values (see below), which may not be true if the two compounds were not accreted or condensed concurrently and/or in the same regions of the solar nebula. In that case, the trend of values from Equation A2 should be seen as more of a proof of concept rather than an accurate model of ice mixtures in comets. Cognizant of this limitation, we may consider mixing to reflect a situation where some interstellar materials underwent chemical processing at closer heliocentric distances and were transported outward in the protoplanetary disk (e.g., Brownlee et al., 2006), while other interstellar materials may have experienced minimal processing prior to accretion (e.g., Drozdovskaya et al., 2021). We can use Equation A2 to explore how the fractionation factor could change as a function of the observed D/H ratio in water constrained by cometary measurements (see Section 3.1). The D/H ratio for the solar source is 2.2×10$^{-5}$ (Geiss and Gloeckler, 2003; Aléon et al., 2022), and a common assumption for the D/H ratio of water ice inherited from an interstellar source is 1×10$^{-3}$ (Alexander et al., 2017a).

To use Equation A2, we need a value for the D/H ratio of interstellar methane. While model predictions (~10$^{-3}$-10$^{-2}$) have been made (e.g., Willacy, 2007; Albertsson et al., 2013), we believe that an approach that is more applicable to the solar system is to estimate this value by applying our mixing model to comet 67P. We thus estimate the interstellar fraction based on the D/H ratio of water in comet 67P, using the following equation derived from Equation A1

$$f_{int} = \frac{R_{H_2O,obs} - R_{sol}}{R_{H_2O,int} - R_{sol}},$$ (A3)

where the 67P value of $R_{H2O,obs}$ is (5.01±0.40)×10$^{-4}$ (Müller et al., 2022). Solving Equation A3, we find that $f_{int}$ should be between ~0.45 and ~0.53, consistent with Alexander et al.'s (2017a) value of 0.52. We next solve for the D/H ratio of interstellar methane that would have been incorporated into comet 67P. We base this calculation on the D/H ratio of 67P methane: $R_{CH_4,obs}$ = (2.41±0.29)×10$^{-3}$ (Müller et al., 2022), and use this value along with the previously determined $f_{int}$ range to evaluate the following expression:

$$R_{CH_4,int} = \frac{R_{CH_4,obs} - (1-f_{int})R_{sol}}{f_{int}}.$$ (A4)

We find $R_{CH4,int}$ = (5±1)×10$^{-3}$. This value falls within the range of D/H ratios predicted by astrochemical models of dark molecular clouds (see above). Given that such consistency is obtained using a mixing model in which methane and water are assumed to have the same $f_{int}$ values, it may be suggested that this assumption is perhaps appropriate, at least as a rough approximation. With a value for $R_{CH4,int}$ in hand, we now have a model that can provide a physically consistent representation of the empirical fractionation factor.

Next, we evaluate Equation A2 to test whether the assumption of constant $\alpha_{CH4-H2O}^{OSN}$ is reasonable. The results of this evaluation are shown in Figure A1. It can be seen that our adopted $\alpha_{CH4-H2O}^{OSN}$ is mostly representative, although $\alpha_{CH4-H2O}^{OSN}$ can reach lower values than the range in Table 1. The lower limit for the mixing model is ~3.6 at $R_{H2O,obs}$ = 1.37×10$^{-4}$. The fractionation factor can reach much lower values, even down to unity, but only outside the cometary range that is considered most relevant to this work (see Section 3.1). That is where the influence of solar methane starts to dominate.

**Figure A1.** D/H fractionation factor between methane and water in the outer solar nebula from a mixing model (yellow region) of interstellar and solar methane, compared to our canonical model (blue region) in which $\alpha_{CH4-H2O}^{OSN}$ is assumed to have a constant value equal to the comet 67P value. Black curves show limiting values for the D/H ratio of an interstellar source that would be consistent with comet 67P data (see Appendix text). The star indicates the solar endmember. Our current best estimate for the D/H ratio of water on Eris and Makemake is within the range encompassed by Jupiter-family comets (JFCs).

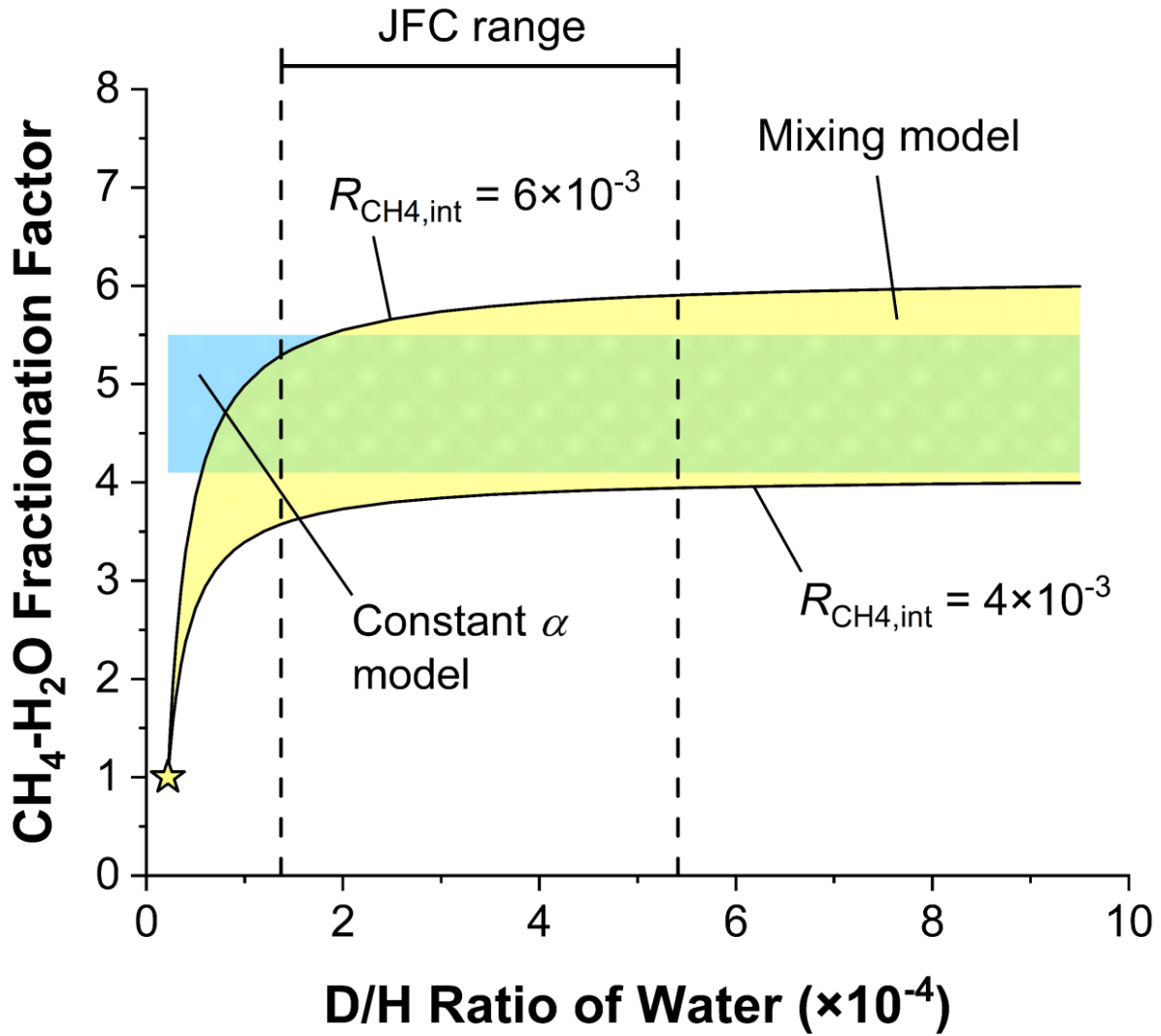

The reason why the assumption of constant $\alpha_{CH4-H2O}^{OSN}$ is largely consistent with results from the mixing model can be deduced by writing an approximate form of Equation A2

$$\alpha_{CH_4-H_2O}^{OSN} \approx \frac{R_{sol}}{R_{H_2O,obs}} + \frac{R_{CH_4,int}}{R_{H_2O,int}} \times \left(1 - \frac{R_{sol}}{R_{H_2O,obs}}\right) \rightarrow \frac{R_{CH_4,int}}{R_{H_2O,int}} = \alpha_{CH_4-H_2O}^{int} \text{ (if } R_{H_2O,obs} \gg R_{sol}\text{),} \quad (A5)$$

which is first simplified by assuming that the solar D/H ratio is negligible compared with the D/H ratios of interstellar methane and water. The equation then approaches the interstellar methane-water fractionation factor when the mixture's D/H ratio in water becomes sufficiently large. "Sufficiently large" is not particularly large in an absolute sense, however. Numerical analysis shows that the mixture's fractionation factor stays within ~20% of the endmember value (i.e., $\alpha_{CH4-H2O}^{int}$) down to $R_{H2O,obs} = 1\times10^{-4}$.

Because the fractionation factor can be lower than we previously assumed, it is necessary to consider how much this could change predicted D/H ratios for primordial methane. Figure A2 shows values obtained by evaluating Equations 2, 3, and A2. It compares these values to those for the constant $\alpha_{CH4-H2O}^{OSN}$ model. There is agreement over the cometary range of interest. Although the mixing model with variable $\alpha_{CH4-H2O}^{OSN}$ can provide methane with lower D/H ratios, they are apparently not low enough to explain the data from Eris and Makemake (Figure A2). Nevertheless, these results suggest that a more conservative lower limit for the D/H ratio of primordial methane in the protoplanetary disk region where TNOs formed is $4.9\times10^{-4}$.

**Figure A2.** Predicted D/H ratios for primordial methane based on a model with constant $\alpha_{CH4-H2O}^{OSN}$ (darker blue region), or based on a mixing model (yellow region outlined in black) that has a variable fractionation factor (see Figure A1) depending on the proportions of interstellar and solar methane that constitute primordial methane. Our current best estimate for the D/H ratio of water on Eris and Makemake is within the range encompassed by Jupiter-family comets (JFCs). The total 1σ range for Eris/Makemake is also shown.

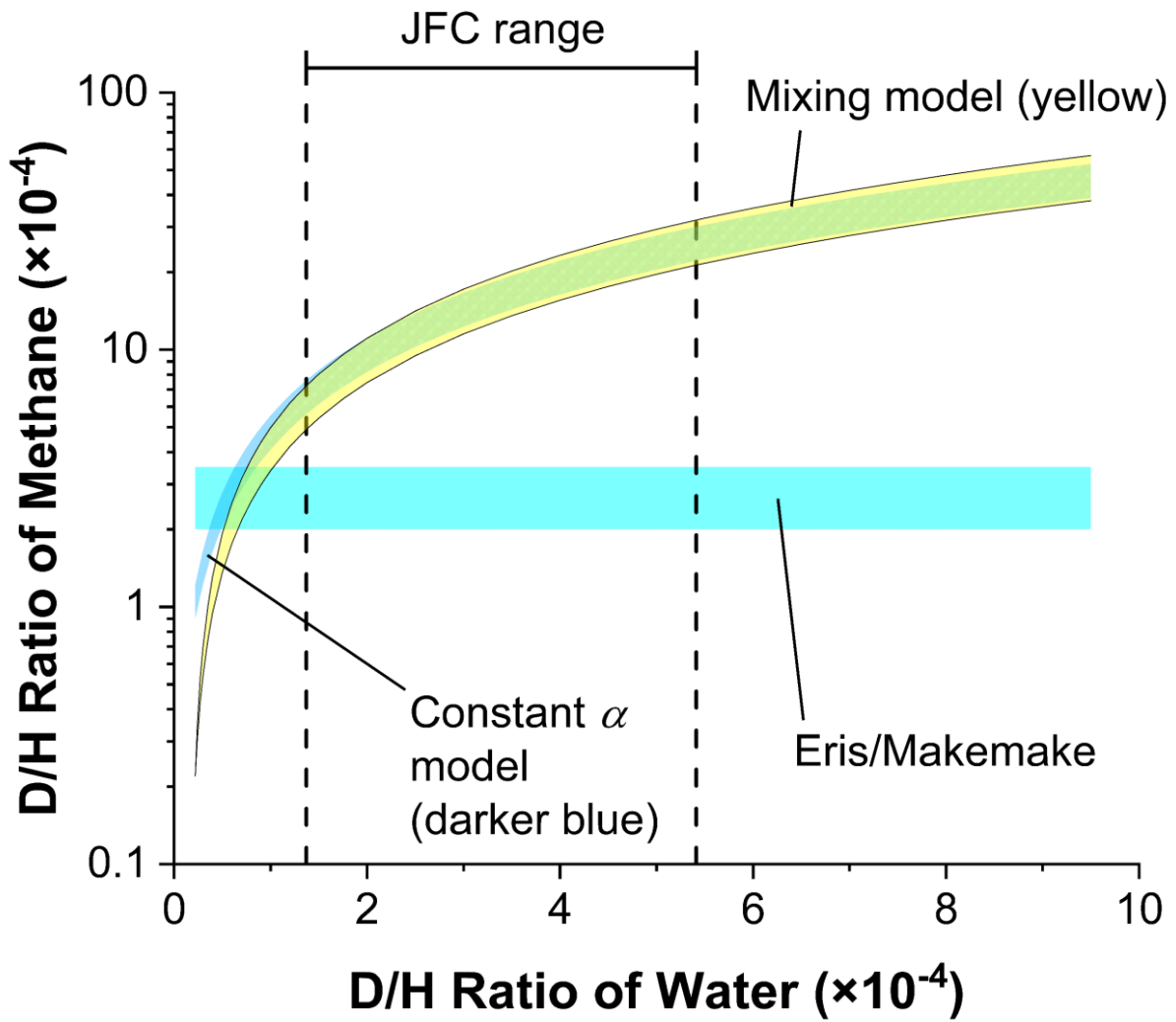